\documentclass[10pt,a4paper]{article}
\usepackage[numbers,sort&compress]{natbib}
\usepackage{authblk}
\usepackage{graphicx}
\usepackage[top=30truemm,bottom=30truemm,left=25truemm,right=25truemm]{geometry}
\usepackage[colorlinks=true,linktoc=page,citecolor=red,linkcolor=blue]{hyperref}
\graphicspath{{./figs/}}
\usepackage{color,comment}
\usepackage{amsmath,amssymb,amsfonts}
\usepackage{bm,braket,array}
\usepackage[all]{xy}
\usepackage{amsthm,amscd}
\usepackage{slashed}
\usepackage[format=hang,justification=raggedright]{caption}
\usepackage{multirow}
\usepackage{rotating}

\newtheorem{thm}{Theorem}[section]

\newtheorem{lem}[thm]{Lemma}

\theoremstyle{definition}
\newtheorem{dfn}[thm]{Definition}
\theoremstyle{remark}

\newcommand{\im}{{\rm im\,}}

\newcommand{\tr}{{\rm tr\,}}

\newcommand{\s}{{\sigma}}

\newcommand{\G}{{\Gamma}}
\newcommand{\C}{\mathbb{C}}

\newcommand{\R}{\mathbb{R}}
\newcommand{\Z}{\mathbb{Z}}

\newcommand{\bk}{{\bm{k}}}
\newcommand{\br}{{\bm{r}}}

\def\widebar{\accentset{{\cc@style\underline{\mskip10mu}}}}
\def\wideubar{\underaccent{{\cc@style\underline{\mskip10mu}}}}

\begin{document}
\title{Intrinsic non-Hermitian topological phases}
\author{Ken Shiozaki}
\affil{Center for Gravitational Physics and Quantum Information, Yukawa Institute for Theoretical Physics, Kyoto University, Kyoto 606-8502, Japan}
\date{\today}
\maketitle

\begin{abstract}
We study the interplay of non-Hermitian topological phases under point- and line-gap conditions.  
Using natural homomorphisms from line-gap to point-gap phases, we distinguish extrinsic phases, reducible to Hermitian or anti-Hermitian line-gapped phases, from intrinsic phases, which are genuinely non-Hermitian without Hermitian counterparts.  
Although classification tables for all symmetry classes were already presented in earlier work, the present paper develops a unified formulation and provides explicit computations for all internal symmetries.  
\end{abstract}

\section{Introduction}

Non-Hermitian quantum systems have recently attracted broad interest across condensed matter physics, open quantum systems, photonic and acoustic platforms, and mechanical metamaterials, providing a setting for novel phenomena absent in Hermitian systems (see reviews~\cite{AshidaNonHermitian2020,BergholtzExceptional2021,OkumaNonHermitian2023,LinTopological2023,WangNonHermitian2023}).  
Compared with Hermitian systems, non-Hermitian systems exhibit several distinctive features: eigenvalues are generally complex, right and left eigenvectors differ, and Hamiltonians may fail to be diagonalizable, giving rise to exceptional points.  
Moreover, spectral properties often show strong boundary dependence, which is called the non-Hermitian skin effect~\cite{YaoEdge2018}, where bulk spectra under periodic and open boundary conditions differ drastically and an extensive number of eigenstates accumulate at the boundary. In such cases, the bulk-boundary correspondence must be reformulated in terms of the generalized Brillouin zone~\cite{YaoEdge2018,YokomizoNonBloch2019}.  

These intrinsic properties of non-Hermitian systems require new topological principles, motivating systematic studies of gap conditions and classification schemes for non-Hermitian Hamiltonians.  
One approach to non-Hermitian topology is based on the \emph{point-gap} condition, where the spectrum avoids a reference point in the complex plane~\cite{GongTopological2018}. 
The point gap gives rise to non-Hermitian-specific topological invariants, such as spectral winding numbers. 
Alternatively, \emph{line gaps} with respect to the real or imaginary axis~\cite{KawabataSymmetry2019} relate non-Hermitian phases to Hermitian topological insulators and superconductors, thereby connecting non-Hermitian classifications to those for Altland–Zirnbauer (AZ) classes~\cite{AltlandNonstandard1997,SchnyderClassification2008,KitaevPeriodic2009,RyuTopological2010}.  

This naturally raises the question: what phenomena are genuinely intrinsic to non-Hermitian systems, i.e., not realizable within Hermitian settings?  
For instance, gapless edge states of a Chern insulator remain stable under non-Hermitian perturbations, indicating their essentially Hermitian origin. 
By contrast, the non-Hermitian skin effect emerges from the spectral winding number that vanishes in Hermitian systems and therefore represents a genuine non-Hermitian topological phenomenon~\cite{ZhangCorrespondence2020,OkumaTopological2020}.  

In this work, we propose a systematic framework to classify such intrinsic non-Hermitian topological phases by subtracting line-gap phases from point-gap phases. Since every line-gapped Hamiltonian is also point-gapped, this defines natural homomorphisms from line-gap to point-gap classifications. We identify the image of these maps with extrinsic phases, while point-gap phases outside the image are defined as intrinsic non-Hermitian topological phases.  

Explicit examples of intrinsic non-Hermitian phases have been reported in various dimensions and symmetry classes: one- and two-dimensional $\mathbb{Z}_2$ skin effects with a reciprocal symmetry~\cite{OkumaTopological2020}, an isolated exceptional point appearing on the surface~\cite{DennerExceptional2021} and magnetic field induced non-Hermitian skin effect in three dimensions~\cite{KawabataTopological2021,BesshoNielsenNinomiya2021}, boundary states associated with PT symmetry breaking trajectories~\cite{DennerInfernal2023,NakamuraBulkBoundary2024}, and momentum-space defect structures~\cite{KawabataClassification2019}. 
Systematic analyses of intrinsic non-Hermitian phases are also discussed in~\cite{NakamuraBulkBoundary2024,DennerInfernal2023}.  
As a related development, the recently established detachability of gapless boundary states in Hermitian systems~\cite{AltlandFragility2024} has been interpreted as an instance of imaginary-line-gap extrinsic phases~\cite{NakamuraNonHermitian2025,ShiozakiTheory2025}.

In this paper, we present a unified formalism that incorporates crystalline symmetries and provides explicit computations for all internal symmetry classes. In particular, we obtain classification results for all 38 independent non-Hermitian symmetry classes, distinguishing extrinsic from intrinsic phases (see Tables~\ref{tab:SFH_AZ}--\ref{tab:SFH_quotient_AZ_add}).  
While the classification results already appeared in Supplemental Material of Ref.~\cite{OkumaTopological2020}, the present work develops a comprehensive formulation together with detailed computations for internal symmetries.

The remainder of this paper is organized as follows. 
In Sec.~2 we review basic ingredients of non-Hermitian free-fermion systems, including the definition of symmetry, the Hermitization procedure, and gap conditions. 
In Sec.~3 we introduce point-gap and line-gap topological phases and formulate the natural homomorphisms from line-gap to point-gap classifications.
Sec.~4 presents explicit computations of these homomorphisms for all internal symmetry classes and provides the resulting classification tables. 
We summarize our findings and discuss future directions in Sec.~5.

\section{Some generalities}
\label{sec:gene}
In this section, we review several basic ingredients of non-Hermitian free-fermion systems: the definition of symmetry, the Hermitization procedure, and gap conditions.

\subsection{Symmetry}
We begin with symmetries of non-Hermitian matrices $H_\bk$ defined over a general parameter space $X$, with $\bk \in X$.
Let $G$ be a finite group, and denote its left action on $X$ by
$G \times X \ni (g,\bk) \mapsto g\bk \in X$.
In a non-Hermitian setting, the Hamiltonian is not necessarily Hermitian, and symmetry transformations may involve complex conjugation, transpose, or their combination, the Hermitian conjugate.
To distinguish these possibilities, we introduce two homomorphisms
\begin{align}
\phi,\kappa: G \to \Z/2 = \{\pm 1\},
\end{align}
and adopt the following notation for a complex number or a matrix $M$:
\begin{align}
M^{\phi_g} =
\begin{cases}
M & (\phi_g=1), \\
M^* & (\phi_g=-1),
\end{cases}
\end{align}
\begin{align}
M^{\phi_g,\kappa_g} =
\begin{cases}
M & (\phi_g=1,\kappa_g=1), \\
M^* & (\phi_g=-1,\kappa_g=1), \\
M^\top & (\phi_g=-1,\kappa_g=-1), \\
M^\dag & (\phi_g=1,\kappa_g=-1).
\end{cases}
\end{align}

For a family of non-Hermitian matrices on $X$, we consider unitary matrices $u_\bk(g)$ and $v_\bk(g)$, defined continuously on $X$, that satisfy the symmetry constraint
\begin{align}
u_\bk(g)\, H_\bk^{\phi_g,\kappa_g}\, v_\bk(g)^\dag = H_{g\bk}, \quad g \in G.
\label{eq:def_sym}
\end{align}
Note that, unlike in the Hermitian case, it is not required that $u_\bk(g)=v_\bk(g)$.  
For $g,h \in G$, comparing two decompositions,
\begin{align}
H_{gh\bk} 
&= u_\bk(gh) H_\bk^{\phi_{gh},\kappa_{gh}} v_\bk(gh)^\dag \nonumber \\
&=
\begin{cases}
u_{h\bk}(g)\, u_\bk(h)^{\phi_g}\, H_\bk^{\phi_{gh},\kappa_{gh}} \left(v_{h\bk}(g)\, v_\bk(h)^{\phi_g}\right)^\dag & (\kappa_g=1), \\
u_{h\bk}(g)\, v_\bk(h)^{\phi_g}\, H_\bk^{\phi_{gh},\kappa_{gh}} \left(v_{h\bk}(g)\, u_\bk(h)^{\phi_g}\right)^\dag & (\kappa_g=-1),
\end{cases}
\end{align}
we obtain
\begin{align}
\begin{cases}
u_\bk(gh)^\dag\, u_{h\bk}(g)\, u_\bk(h)^{\phi_g}\, H_\bk^{\phi_{gh},\kappa_{gh}}
= H_\bk^{\phi_{gh},\kappa_{gh}}\, v_\bk(gh)^\dag\, v_{h\bk}(g)\, v_\bk(h)^{\phi_g} & (\kappa_g=1), \\
u_\bk(gh)^\dag\, u_{h\bk}(g)\, v_\bk(h)^{\phi_g}\, H_\bk^{\phi_{gh},\kappa_{gh}}
= H_\bk^{\phi_{gh},\kappa_{gh}}\, v_\bk(gh)^\dag\, v_{h\bk}(g)\, u_\bk(h)^{\phi_g} & (\kappa_g=-1).
\end{cases}
\label{eq:gh_vs_g_h}
\end{align}
We assume that the symmetry group $G$ is chosen appropriately such that \eqref{eq:gh_vs_g_h} imposes only trivial constraints~\footnote{If the product $u_\bk(gh)^\dag u_{h\bk}(g) u_\bk(h)^{phi_g}$ (and similarly for $v_\bk(g)$) were not a $U(1)$ phase, then the symmetry would not be described by the group $G$, but rather by a group $\tilde G$ that is a nontrivial extension of $G$ by another group as $u_\bk(g)$ is not a projective representation of $G$}.
In other words, we suppose there exists a $U(1)$ phase $z_\bk(g,h)$ such that
\begin{align}
\begin{cases}
u_\bk(gh)^\dag\, u_{h\bk}(g)\, u_\bk(h)^{\phi_g}
= \left(v_\bk(gh)^\dag\, v_{h\bk}(g)\, v_\bk(h)^{\phi_g}\right)^\dag
= z_\bk(g,h) & (\kappa_g=1), \\
u_\bk(gh)^\dag\, u_{h\bk}(g)\, v_\bk(h)^{\phi_g}
= \left(v_\bk(gh)^\dag\, v_{h\bk}(g)\, u_\bk(h)^{\phi_g}\right)^\dag
= z_\bk(g,h) & (\kappa_g=-1),
\end{cases}
\label{eq:uv_assoc}
\end{align}
holds.  
The $U(1)$ phase $z_\bk(g,h)$ must then satisfy the following cocycle condition depending on $\bk \in X$:
\begin{align}
z_{h,l}(\bk)^{\phi_g}\, z_{gh,l}(\bk)^{-1}\, z_{g,hl}(\bk)\, z_{g,h}(l\bk)^{-1}=1,
\quad g,h,l \in G,\ \bk \in X.
\label{eq:z_cocycle}
\end{align}

\subsection{Hermitization}
For a non-Hermitian Hamiltonian $H_\bk$, we double the matrix size and define the Hermitian Hamiltonian
\begin{align}
\tilde H_\bk := \begin{pmatrix}
& H_\bk \\
H_\bk^\dag & 
\end{pmatrix}_\s ,
\label{eq:doubled_ham}
\end{align}
where the subscript $\sigma$ indicates the block structure associated with the doubled internal degrees of freedom.  
We denote by $\s_\mu$ $(\mu=0,x,y,z)$ the Pauli matrices acting on this doubled space.  
The corresponding doubled symmetry operators are defined by
\begin{align}
\tilde u_\bk(g) := 
\begin{cases}
\begin{pmatrix}
u_\bk(g) & \\
& v_\bk(g) \\
\end{pmatrix} & (\kappa_g=1), \\[1.2ex]
\begin{pmatrix}
& u_\bk(g) \\
v_\bk(g) & \\
\end{pmatrix} & (\kappa_g=-1),
\end{cases}
\quad g \in G,
\label{eq:tilde_u}
\end{align}
so that the symmetry relations (\ref{eq:def_sym}) and (\ref{eq:uv_assoc}) can be written more compactly in the Hermitianized form:
\begin{align}
&\tilde u_\bk(g)\, \tilde H_\bk^{\phi_g}\, \tilde u_\bk(g)^\dag = \tilde H_{g\bk}, \quad g \in G, \nonumber \\
&\tilde u_{h\bk}(g)\, \tilde u_\bk(h)^{\phi_g} = z_\bk(g,h)\, \tilde u_\bk(gh), \quad g,h \in G.
\label{eq:H_G_sym}
\end{align}
The doubling in (\ref{eq:doubled_ham}) and (\ref{eq:tilde_u}) also implies the chiral symmetry
\begin{align}
\s_z \tilde H_\bk \s_z = - \tilde H_\bk ,
\label{eq:H_chi_sym}
\end{align}
together with the algebraic relation between $\tilde u_\bk(g)$ and the chiral operator,
\begin{align}
\s_z \tilde u_\bk(g) = \kappa_g\, \tilde u_\bk(g) \s_z, \quad g \in G.
\label{eq:H_u_chi}
\end{align}

Conversely, the conditions (\ref{eq:H_G_sym}), (\ref{eq:H_chi_sym}), and (\ref{eq:H_u_chi}) imply the original non-Hermitian symmetry relation (\ref{eq:def_sym}).  
Thus, the topological properties of a non-Hermitian Hamiltonian $H_\bk$ can be studied equivalently through its Hermitianized counterpart $\tilde H_\bk$.

\subsection{Point-gap condition}
For a family of non-Hermitian Hamiltonians $H_\bk$, we would like to define a notion of topologically stable classification.  
Since any two Hamiltonians $H_\bk,H_\bk'$ can be linearly interpolated while preserving the symmetry (\ref{eq:def_sym}) as
\begin{align}
(1-t) H_\bk + t H_\bk', \quad t \in [0,1],
\end{align}
no nontrivial classification can be obtained without an additional constraint.  
To obtain a meaningful classification, we impose gap conditions as described below.

\subsubsection{Singular-value gap}
\label{sec:s_gap}
We can regard a non-Hermitian Hamiltonian $H_\bk$ as a family of linear maps
$H_\bk: {\cal H}_1 \to {\cal H}_2$ acting between two $N$-dimensional Hilbert spaces 
${\cal H}_1 \cong {\cal H}_2 \cong \C^N$.  
Under independent basis changes $U_1,U_2$ of ${\cal H}_1,{\cal H}_2$, the Hamiltonian transforms as
\begin{align}
H_\bk \mapsto U_1^\dag H_\bk U_2 .
\label{eq:Hk_basis_change}
\end{align}
Note that the set of singular values
\begin{align}
\s(H_\bk) = \left\{ \lambda \in \R_{\geq 0} \,\big|\, \det \!\left(\lambda^2 {\bf 1}_N - H_\bk^\dag H_\bk \right)=0 \right\}
\end{align}
is invariant under the transformation (\ref{eq:Hk_basis_change}).  
If ${\cal H}_1$ and ${\cal H}_2$ are regarded as distinct physical spaces, it is natural to introduce a gap condition that is invariant under such basis changes.  
To this end, consider a subset of the nonnegative real axis given by a single interval
$I=[a,b]$, $0 \leq a < b$, or by a disjoint union of intervals
$I = [a,b] \cup [c,d] \cup \cdots$, with $0 \leq a<b<c<d<\cdots$.  
We then impose the singular-value gap condition
\begin{align}
\left( \bigcup_{\bk \in X} \sigma(H_\bk) \right) \cap I = \emptyset .
\end{align}
As a special case, taking $I=[0,\epsilon]$ with some $\epsilon>0$, or equivalently requiring
\begin{align}
\det H_\bk \neq 0 \quad \text{for all }\bk \in X,
\end{align}
is called the point-gap condition~\cite{GongTopological2018}.  

Since the condition $\det H_\bk=0$ is equivalent to $\det \tilde H_\bk=0$, the point-gap condition for a non-Hermitian Hamiltonian $H_\bk$ is equivalent to the usual gap condition for the Hermitianized Hamiltonian $\tilde H_\bk$, i.e., that $\tilde H_\bk$ has no zero eigenvalues.

\subsubsection{Constraint of symmetry matrices}
If a non-Hermitian Hamiltonian $H_\bk$ satisfies the point-gap condition, then it is invertible for all $\bk \in X$.  
This imposes strong restrictions on the possible symmetry matrices $u_\bk(g), v_\bk(g)$.  
In fact~\cite{ShiozakiSymmetry2021},
\begin{align}
\text{there exists a continuous family of invertible matrices $M_\bk$ such that\ }
v_\bk(g) = M_{g\bk}^{-1}\, u_\bk(g)\, M_\bk^{\phi_g,\kappa_g}.
\label{eq:u_v_rel}
\end{align}
(The Hamiltonian $H_\bk$ itself provides an example of such a family $M_\bk$.)  
In particular, for the subgroup $G_\bk \cap \ker \kappa = \{ g \in G \mid g\bk=\bk,\ \kappa_g=1 \}$, this relation means that $u_\bk(g)$ and $v_\bk(g)$ are unitarily equivalent at fixed points $g\bk = \bk$. 
Conversely, if (\ref{eq:u_v_rel}) fails to hold for the symmetry matrices, then $H_\bk$ never satisfies the point-gap condition over all of $X$, and the gap must close at some point $\bk \in X$.  

A similar argument applies to Hermitian systems independently of non-Hermitian ones.  
Namely, if a Hermitian Hamiltonian $\tilde H_\bk$ satisfies the symmetries (\ref{eq:H_G_sym}), (\ref{eq:H_chi_sym}), (\ref{eq:H_u_chi}) and has no zero eigenvalues for all $\bk \in X$, then the symmetry operators $u_\bk(g),v_\bk(g)$ defined by (\ref{eq:tilde_u}) necessarily satisfy (\ref{eq:u_v_rel}).  
Conversely, if the symmetry operators fail to satisfy (\ref{eq:u_v_rel}), the Hermitian Hamiltonian $\tilde H_\bk$ must become gapless somewhere.

\subsubsection{Choice of symmetry matrices}
In this work, we focus on genuinely non-Hermitian phases with point gaps that cannot be understood as Hermitian or anti-Hermitian systems.  
Therefore, we always impose the point-gap condition, and the possible symmetry matrices $u_\bk(g),v_\bk(g)$ cannot be chosen arbitrarily.  

Without pursuing full generality in this paper, we fix the choice as follows.  
For each group element $g \in G$, introduce a homomorphism
\begin{align}
c: G \to \{\pm 1\},
\end{align}
and take
\begin{align}
v_\bk(g) = c_g\, u_\bk(g), \quad g \in G.
\label{eq:choice_of_v}
\end{align}
Then the definition of the symmetry (\ref{eq:def_sym}) becomes
\begin{align}
u_\bk(g)\, H_\bk^{\phi_g,\kappa_g}\, u_\bk(g)^\dag = c_g\, H_{g\bk}, \quad g \in G.
\label{eq_def_sym_2}
\end{align}
Apart from the conjugations specified by $\kappa_g$, this is the standard form of symmetries in Hermitian systems.  
Each element $g \in G$ is thus characterized by the three homomorphisms $\phi,c,\kappa: G \to \Z_2$, giving eight possible types in total.  
Their names are summarized in Table~\ref{tab:sym_type}~\cite{KawabataSymmetry2019}.

\begin{table}[!]
\begin{center}
\caption{Eight possible types of symmetries.}
\label{tab:sym_type}
\begin{tabular}{ccccccc}
$\phi$ & $c$ & $\kappa$ & Type & Abbreviation \\
\hline
1 & 1 & 1 & Unitary & Uni \\
-1 & 1 & 1 & Time-reversal symmetry & TRS \\
-1 & -1 & -1 & Particle-hole symmetry & PHS \\
1 & -1 & -1 & Chiral symmetry & CS \\
1 & 1 & -1 & Pseudo-Hermiticity & pH \\
-1 & 1 & -1 & Time-reversal symmetry dagger & TRS$^\dag$ \\
-1 & -1 & 1 & Particle-hole symmetry dagger & PHS$^\dag$ \\
1 & -1 & 1 & Sublattice symmetry & SLS \\
\end{tabular}
\end{center}
\end{table}

The choice (\ref{eq:choice_of_v}) is not fully general even under the point-gap condition.  
For example, for a group element $\s \in G$ of order $n$, $\s^n=e$, one may consider
\begin{align}
v_\bk(\s) = e^{\frac{2\pi i p}{n}}\, u_\bk(\s), \quad p=0,\dots,n-1.
\label{eq:choice_v_order_n}
\end{align}
If $n \geq 3$ and $p \geq 3$, such choices are not compatible with the line-gap conditions introduced below (real-line and imaginary-line gaps).  
In that case, the notion of line gaps must be generalized (which we do not discuss here).  

Finally, note that when $\kappa_g=-1$, the restriction (\ref{eq:u_v_rel}) does not imply unitary equivalence between $u_\bk(g)$ and $v_\bk(g)$.  
Thus, for $\kappa_g=-1$, the possible forms of $v_\bk(g)$ under the point-gap condition have additional freedom beyond (\ref{eq_def_sym_2}).  
See Ref.~\cite{ShiozakiSymmetry2021} for explicit examples.

\subsection{Line gaps}
Introducing line-gap conditions allows us to distinguish, among point-gapped topological phases, which phases can be understood in terms of Hermitian systems.  
The symmetry relation (\ref{eq_def_sym_2}) relates eigenvalues in quartets $(E, E^*, -E, -E^*)$.  
It is therefore natural to consider the following two types of line gaps~\cite{KawabataSymmetry2019}.
The spectrum of a matrix $M$ is denoted by
${\rm Sp}(M) = \{\lambda \in \mathbb{C} \;|\; \det(\lambda {\bf 1}_N - M) = 0\}$.

\paragraph{Real line gap.}
We say that $H_\bk$ has a real line gap if its spectrum avoids the imaginary axis, namely,
\begin{align}
\left(\bigcup_{\bk \in X} {\rm Sp}(H_\bk)\right) \cap i\R = \emptyset.
\end{align}

\paragraph{Imaginary line gap.}
We say that $H_\bk$ has an imaginary line gap if its spectrum avoids the real axis, namely,
\begin{align}
\left(\bigcup_{\bk \in X} {\rm Sp}(H_\bk)\right) \cap \R = \emptyset.
\end{align}

Note that in the above we chose the imaginary axis $i\R$ and the real axis $\R$ as the reference lines.  
In the absence of symmetries relating all of $E$, $-E$, $E^*$, and $-E^*$, the reference line can in general be shifted.  
For example, if only TRS-type symmetry is present, then the spectrum is symmetric with respect to the real axis.  
In this case there is no distinguished reference point along the real direction, and one may instead take $a+i\R=\{\,a+iy \mid y\in\R\,\}$ as the reference line for the real line gap, with any $a\in\R$.

An important fact is that: 
\begin{lem}[(Anti-)Hermitization]
\label{lem:linegap}
If a Hamiltonian $H_\bk$ has a real line gap, then it can be continuously deformed, while preserving both the real line gap and the symmetry (\ref{eq_def_sym_2}), into a Hermitian Hamiltonian $H'_\bk$ with $H'_\bk=H'_\bk{}^\dag$.  
Similarly, if $H_\bk$ has an imaginary line gap, then it can be continuously deformed, while preserving both the imaginary line gap and the symmetry (\ref{eq_def_sym_2}), into an anti-Hermitian Hamiltonian $H'_\bk$ with $H'_\bk{}^\dag=-H'_\bk$.
\end{lem}

The proof follows the argument in Appendix D of Ref.~\cite{AshidaNonHermitian2020}, so we only outline it here.  
If $H_\bk$ has a real line gap, introduce closed contours $C_\pm$ encircling all eigenvalues with $\Re(E_\bk)>0$ and $\Re(E_\bk)<0$, respectively.  
Define the projectors $P^{\pm}_\bk := \oint_{C_\pm} \frac{dz}{2\pi i} \frac{1}{z-H_\bk}$. 
The flattened Hamiltonian $Q_\bk = P^+_\bk - P^-_\bk$ has eigenvalues $\pm 1$ and preserves the symmetry (\ref{eq_def_sym_2})\footnote{Because the symmetry operators satisfy $v_\bk(g)=\pm u_\bk(g)$, the group action $H \mapsto u_\bk(g)\, H_\bk^{\phi_g,\kappa_g}\, v_\bk(g)^\dag$ extends, up to a sign, to any matrix function $f(H_\bk)$. In particular, one finds $u_\bk(g) \left(\frac{1}{z-H_\bk}\right)^{\phi_g,\kappa_g} v_\bk(g)^\dag = \frac{1}{z\mp H_{g\bk}}$. If $v_\bk(g)=\pm u_\bk(g)$ does not hold, then it is not guaranteed that one can Hermitize (or anti-Hermitize) while preserving both the line gap and the symmetry~\cite{ShiozakiSymmetry2021}.}.  
Decompose $Q_\bk = H'_\bk + i H''_\bk$, where $H'_\bk = \frac{Q_\bk+Q_\bk^\dag}{2}, H''_\bk = \frac{Q_\bk-Q_\bk^\dag}{2i}$. The linear interpolation $(1-t)Q_\bk+tH'_\bk$, $t\in[0,1]$, preserves both the real line gap and the symmetry (\ref{eq_def_sym_2}).  
For the imaginary line gap, the same construction applies after replacing $H_\bk \mapsto i H_\bk$.  

\section{Intrinsic and extrinsic non-Hermitian topological phases}
In this section, we define point-gap topological phases as well as real- and imaginary-line-gap topological phases, and formulate homomorphisms from line-gap phases to point-gap phases.

\subsection{Symmetry}
Based on Sec.~\ref{sec:gene}, we now re-summarize the symmetry structure adopted for non-Hermitian systems in this work.

Let $X$ be the parameter space, $G$ a finite group, and write $g\bk$ for the left $G$-action on $X$.  
Let $\phi,c,\kappa: G \to \{\pm 1\}$ be homomorphisms, and let $z_\bk(g,h)$ be a 2-cocycle satisfying (\ref{eq:z_cocycle}).  
The $G$-symmetry of a Hamiltonian $H_\bk$ is defined by
\begin{align}
\begin{cases}
u_\bk(g)\, H_\bk^{\phi_g,\kappa_g}\, u_\bk(g)^\dag = c_g H_{g\bk}, \quad g \in G, \\
u_{h\bk}(g)\, u_{\bk}(h)^{\phi_g} = z_\bk(g,h)\, u_\bk(gh), \quad g,h \in G, 
\label{eq:ug_cond}
\end{cases}
\end{align}
where $u_\bk(g)$ are continuous unitary matrices on $X$.  
Within the framework of $K$-theoretic classification, the symmetry matrices $u_\bk(g)$ are not fixed; instead, one considers all possible forms of $u_\bk(g)$ satisfying (\ref{eq:ug_cond}) for a given 2-cocycle $z_\bk(g,h)$.  
Thus, the data specifying a non-Hermitian system consist of the finite group $G$, the parameter space $X$, the left action
\begin{align}
\alpha: G \times X \to X, \quad (g,\bk) \mapsto g\bk,
\end{align}
the homomorphisms $\phi,c,\kappa : G \to \{\pm 1\}$, and the 2-cocycle 
$z \in Z^2_{\rm group}(G,C(X,U(1)_\phi))$, where
\begin{align}
(G,X,\alpha,\phi,c,\kappa,z)
\end{align}
collectively define a non-Hermitian system~\footnote{The group cocycle is defined as follows.  
Let $C(X,U(1)_\phi)$ denote the abelian group of continuous $U(1)$-valued functions $f:X\to U(1)$, with pointwise multiplication $(ff')_\bk = f_\bk f'_\bk$.  
The left and right actions of $G$ on $C(X,U(1)_\phi)$ are given, for $g \in G$, by
$(g \cdot f)_\bk = f_\bk^{\phi_g}$ and $(f\cdot g)_\bk = f_{g\bk}$.  
Then $z_\bk$ is a 2-cocycle in the group cochain complex $C^*_{\rm group}(G,C(X,U(1)_\phi))$.}.

\subsection{Point-gap and line-gap topological phases}
\label{sec:P_L_top_phase}
As noted in Sec.~\ref{sec:s_gap}, the existence of a point gap in $H_\bk$ is equivalent to the doubled Hermitian Hamiltonian $\tilde H_\bk$ being gapped.  
We can therefore adopt the well-established $K$-theoretic definition of topological phases~\cite{KitaevPeriodic2009,FreedTwisted2013,ThiangKTheoretic2016,ShiozakiTopological2017}.

We redefine the symmetry operators for the doubled Hermitian Hamiltonian \eqref{eq:doubled_ham} as
\begin{align}
U_\bk(g) := u_\bk(g) \otimes (\s_x)^{\frac{1-\kappa_g}{2}}, \quad g \in G. 
\label{eq:U_def}
\end{align}
With this choice, the symmetry of the Hermitianized Hamiltonian $\tilde H_\bk$ can be summarized as follows.

\begin{dfn}[Point-gap topological phase]
The classification group $K_{\rm P}(G,X,\alpha,\phi,c,\kappa,z)$ of point-gap topological phases is defined as the $K$-group of gapped Hermitian Hamiltonians satisfying symmetry constraitns: 
\begin{align}
K_{\rm P}(G,X,\alpha,\phi,c,\kappa,z): \quad 
\begin{cases}
U_\bk(g)\, \tilde H_\bk^{\phi_g}\, U_\bk(g)^\dag = c_g \tilde H_{g\bk}, \quad g \in G, \\
U_{h\bk}(g)\, U_{\bk}(h)^{\phi_g} = z_\bk(g,h)\, U_\bk(gh), \quad g,h \in G, \\
\sigma_z \tilde H_\bk \sigma_z = -\tilde H_\bk, \\
U_\bk(g) \s_z = \kappa_g \s_z U_\bk(g), \quad g \in G.
\end{cases}
\label{eq:sym_KP}
\end{align}
\end{dfn}

Two remarks are in order. 

The original symmetry matrices $u_\bk(g)$ of the non-Hermitian Hamiltonian $H_\bk$ need not be expressible in the form of (\ref{eq:ug_cond}).  
In general, they are defined via the restriction $U_\bk(g)\s_z=\kappa_g \s_z U_\bk(g)$ in (\ref{eq:sym_KP}).  
Thus, the definition (\ref{eq:sym_KP}) encompasses a broader class of symmetries in non-Hermitian systems, including but not limited to those of the form (\ref{eq:ug_cond}).

Using the chiral symmetry $\s_z$, one can reinterpret (\ref{eq:sym_KP}) as the symmetry of a Hermitian Hamiltonian defined over the suspension of the parameter space.  
Define the Hamiltonian on the suspension $S X = X \times [0,\pi]/(X \times \{0\} \cup X \times \{\pi\})$ by
\begin{align}
\tilde H'_{\bk,\theta} := \cos \theta\, \tilde H_\bk + \sin \theta\, \sigma_z .
\end{align}
Then $\tilde H'_{\bk,\theta}$ satisfies the symmetry constraints, and we get an alternative expression of the classification group $K_{\rm P}(G,X,\alpha,\phi,c,\kappa,z)$: 
\begin{align}
K_{\rm P}(G,X,\alpha,\phi,c,\kappa,z): \quad 
\begin{cases}
U_\bk(g)\, (\tilde H'_{\bk,\theta})^{\phi_g}\, U_\bk(g)^\dag = c_g \tilde H'_{g\bk,\,c_g\kappa_g \theta}, \quad g \in G, \\
U_{h\bk}(g)\, U_{\bk}(h)^{\phi_g} = z_\bk(g,h)\, U_\bk(gh), \quad g,h \in G,
\end{cases}
\label{eq:sym_KP_suspension}
\end{align}
with the additional $G$-action on $\theta$ given by $\theta \mapsto c_g \kappa_g \theta$.  
If the $G$-action leaves a base point $\bk_*\in X$ fixed, then by suspension isomorphism the $K$-groups defined by (\ref{eq:sym_KP}) and (\ref{eq:sym_KP_suspension}) agree up to contributions from the base point.

\vspace{0.5em}

We now define the classification groups for real- and imaginary-line-gap topological phases.  
By Lemma~\ref{lem:linegap}, if $H_\bk$ has a real line gap, it can be continuously deformed, while preserving both the real line gap and the symmetry (\ref{eq:ug_cond}), into a Hermitian Hamiltonian with $H_\bk^\dag = H_\bk$.  
For the doubled Hermitianized Hamiltonian $\tilde H_\bk$, the Hermitian condition $H_\bk^\dag=H_\bk$ is equivalent to imposing an additional chiral symmetry $\s_y \tilde H_\bk \s_y = -\tilde H_\bk$, or equivalently an additional $\Z_2$ symmetry $\s_x \tilde H_\bk \s_x = \tilde H_\bk$.

\begin{dfn}[Real-line-gap topological phase]
The classification group $K_{\rm L_r}(G,X,\alpha,\phi,c,\kappa,z)$ of real-line-gap topological phases is defined as the $K$-group of gapped Hermitian Hamiltonians satisfying
\begin{align}
K_{\rm L_r}(G,X,\alpha,\phi,c,\kappa,z): \quad 
\begin{cases}
U_\bk(g)\, \tilde H_\bk^{\phi_g}\, U_\bk(g)^\dag = c_g \tilde H_{g\bk}, \quad g \in G, \\
U_{h\bk}(g)\, U_{\bk}(h)^{\phi_g} = z_\bk(g,h)\, U_\bk(gh), \quad g,h \in G, \\
\sigma_z \tilde H_\bk \sigma_z = -\tilde H_\bk, \\
U_\bk(g) \s_z = \kappa_g \s_z U_\bk(g), \quad g \in G, \\
\sigma_y \tilde H_\bk \sigma_y = -\tilde H_\bk, \\
U_\bk(g) (i\s_y) = \kappa_g (i\s_y) U_\bk(g), \quad g \in G.
\end{cases}
\label{eq:sym_KLr}
\end{align}
\end{dfn}

Indeed, when $\{\s_z,\tilde H_\bk\}=\{\s_y,\tilde H_\bk\}=0$, the doubled Hamiltonian can be written as $\tilde H_\bk = H_\bk \otimes \s_x$, with $H_\bk^\dag=H_\bk$.  
The corresponding symmetry relation for $H_\bk$, determined by (\ref{eq:U_def}), is
\begin{align}
K_{\rm L_r}(G,X,\alpha,\phi,c,\kappa,z): \quad 
\begin{cases}
u_\bk(g)\, H_\bk^{\phi_g}\, u_\bk(g)^\dag = c_g H_{g\bk}, \quad g \in G, \\
u_{h\bk}(g)\, u_{\bk}(h)^{\phi_g} = z_\bk(g,h)\, u_\bk(gh), \quad g,h \in G,
\end{cases}
\label{eq:Lr_equ_cond}
\end{align}
which coincides with the standard symmetry condition in Hermitian systems.

\vspace{0.5em}

Similarly, by Lemma~\ref{lem:linegap}, if $H_\bk$ has an imaginary line gap, it can be continuously deformed, while preserving both the imaginary line gap and the symmetry (\ref{eq:ug_cond}), into an anti-Hermitian Hamiltonian with $H_\bk^\dag = -H_\bk$.  
For the doubled Hermitianized Hamiltonian $\tilde H_\bk$, the anti-Hermitian condition is equivalent to imposing an additional chiral symmetry $\s_x \tilde H_\bk \s_x = -\tilde H_\bk$, or equivalently an additional $\Z_2$ symmetry $\s_y \tilde H_\bk \s_y = \tilde H_\bk$.  
Indeed, when $\{\s_z,\tilde H_\bk\}=\{\s_x,\tilde H_\bk\}=0$, one can write $\tilde H_\bk = h_\bk \otimes \s_y$ with $h_\bk^\dag=-h_\bk$.

\begin{dfn}[Imaginary-line-gap topological phase]
The classification group $K_{\rm L_i}(G,X,\alpha,\phi,c,\kappa,z)$ of imaginary-line-gap topological phases is defined as the $K$-group of gapped Hermitian Hamiltonians satisfying
\begin{align}
K_{\rm L_i}(G,X,\alpha,\phi,c,\kappa,z): \quad 
\begin{cases}
U_\bk(g)\, \tilde H_\bk^{\phi_g}\, U_\bk(g)^\dag = c_g \tilde H_{g\bk}, \quad g \in G, \\
U_{h\bk}(g)\, U_{\bk}(h)^{\phi_g} = z_\bk(g,h)\, U_\bk(gh), \quad g,h \in G, \\
\sigma_z \tilde H_\bk \sigma_z = -\tilde H_\bk, \\
U_\bk(g) \s_z = \kappa_g \s_z U_\bk(g), \quad g \in G, \\
\sigma_x \tilde H_\bk \sigma_x = -\tilde H_\bk, \\
U_\bk(g) \s_x = \s_x U_\bk(g), \quad g \in G.
\end{cases}
\label{eq:sym_KLi}
\end{align}
\end{dfn}

By redefining the Hamiltonian as $H \mapsto iH$, one obtains an alternative symmetry constraint suitable for imaginary-line-gap topological phases:
\begin{align}
K_{\rm L_i}(G,X,\alpha,\phi,c,\kappa,z): \quad 
\begin{cases}
u_\bk(g)\, H_\bk^{\phi_g}\, u_\bk(g)^\dag = \phi_g\, c_g\, \kappa_g\, H_{g\bk}, & g \in G, \\
u_{h\bk}(g)\, u_{\bk}(h)^{\phi_g} = z_\bk(g,h)\, u_\bk(gh), & g,h \in G. 
\end{cases}
\end{align}
It should be noted that chiral-type symmetries ($g \in G$ with $(\phi_g,c_g,\kappa_g)=(1,-1,-1)$) effectively act as unitary-type symmetries.

\subsection{Intrinsic and extrinsic}
It is clear that if a non-Hermitian Hamiltonian $H_\bk$ has either a real or an imaginary line gap, then it also has a point gap.  
Thus, real- and imaginary-line-gap topological phases are always well-defined as point-gap topological phases.  
This gives rise to the natural homomorphisms
\begin{align}
&f_{\rm r}: K_{\rm L_r}(G,X,\alpha,\phi,c,\kappa,z) \to K_{\rm P}(G,X,\alpha,\phi,c,\kappa,z), \\
&f_{\rm i}: K_{\rm L_i}(G,X,\alpha,\phi,c,\kappa,z) \to K_{\rm P}(G,X,\alpha,\phi,c,\kappa,z).
\end{align}
These maps are concretely defined by simply forgetting the additional chiral symmetry.

\begin{dfn}[Homomorphisms from line-gap to point-gap phases]
The map $f_{\rm r}$ is defined by forgetting the chiral symmetry $\sigma_y$ in the relations (\ref{eq:sym_KLr}).  
Similarly, the map $f_{\rm i}$ is defined by forgetting the chiral symmetry $\sigma_x$ in the relations (\ref{eq:sym_KLi}).
\end{dfn}

Explicit computations will be presented in the next section.  

Among point-gap topological phases, those that can be deformed into real-line-gap or imaginary-line-gap topological phases while preserving the point gap are precisely the images of $f_{\rm r}$ and $f_{\rm i}$. 
We refer to these as real-line-gap extrinsic topological phases and imaginary-line-gap extrinsic topological phases, respectively.

\begin{dfn}[Extrinsic topological phases]
Within the point-gap classification group $K_{\rm P}(G,X,\alpha,\phi,c,\kappa,z)$, the subgroup of real-line-gap extrinsic topological phases is
\begin{align}
\im f_{\rm r} = \im[f_{\rm r}: K_{\rm L_r}(G,X,\alpha,\phi,c,\kappa,z) \to K_{\rm P}(G,X,\alpha,\phi,c,\kappa,z)].
\end{align}
Likewise, the subgroup of imaginary-line-gap extrinsic topological phases is
\begin{align}
\im f_{\rm i} = \im[f_{\rm i}: K_{\rm L_i}(G,X,\alpha,\phi,c,\kappa,z) \to K_{\rm P}(G,X,\alpha,\phi,c,\kappa,z)].
\end{align}
We define the line-gap extrinsic topological phases as the sum
\begin{align}
\im f_{\rm r} + \im f_{\rm i}
= \{\,x+y \mid x \in \im f_{\rm r},\ y \in \im f_{\rm i}\,\}.
\end{align}
\end{dfn}

Consequently, point-gap topological phases that do not lie in either image $\im f_{\rm r}$ or $\im f_{\rm i}$ are unique to non-Hermitian systems.  
We call these intrinsic topological phases.

\begin{dfn}[Intrinsic topological phases]
Within $K_{\rm P}(G,X,\alpha,\phi,c,\kappa,z)$, the intrinsic topological phases are in the complement set 
\begin{align}
K_{\rm P}(G,X,\alpha,\phi,c,\kappa,z) \setminus (\im f_{\rm r} + \im f_{\rm i}).
\end{align}
Equivalently, the classification of intrinsic phases is given by the quotient group
\begin{align}
K_{\rm P}(G,X,\alpha,\phi,c,\kappa,z)/(\im f_{\rm r} + \im f_{\rm i}).
\end{align}
\end{dfn}

\subsection{On degree shifts of $K$-groups}
In this subsection, we comment on the relation to the standard notation of twisted equivariant $K$-theory~\cite{FreedTwisted2013,GomiFreedMoore2017}.

In $K$-theory, a (Hermitian) Hamiltonian equipped with a chiral symmetry may be absorbed as a degree shift of the corresponding $K$-group.  
Accordingly, the classification group $K_{\rm P}(G,X,\alpha,\phi,c,\kappa,z)$ for point-gap topological phases can be expressed as a degree shift of a certain symmetry class.  
For the doubled Hermitian Hamiltonian \eqref{eq:doubled_ham}, redefine the symmetry operators \eqref{eq:tilde_u} by
\begin{align}
U'_\bk(g) := u_\bk(g) \otimes (\s_x)^{\frac{1-\kappa_g}{2}} (\s_z)^{\frac{1-c_g\kappa_g}{2}},\quad g \in G. 
\end{align}
With this choice, the symmetry of the Hermitianized Hamiltonian $\tilde H_\bk$ is summarized as
\begin{align}
\begin{cases}
U'_\bk(g)\tilde H_\bk^{\phi_g} U'_\bk(g)^\dag= \kappa_g \tilde H_{g\bk},\quad g \in G, \\
U'_{h\bk}(g)U'_{\bk}(h)^{\phi_g}=(-1)^{\frac{1-c_g\kappa_g}{2} \frac{1-\kappa_h}{2}} z_\bk(g,h)U'_\bk(gh), \quad g,h \in G,\\
\sigma_z \tilde H_\bk \sigma_z=-\tilde H_{\bk}, \\
U'_\bk(g) \s_z = \kappa_g \s_z U'_\bk(g),\quad g\in G.
\end{cases}
\label{eq:sym_def_doubled}
\end{align}
In general, $\sigma_z$ acts as a chiral symmetry that lowers the $K$-theory degree by $1$ (for example, see \cite{ShiozakiTopological2017}).
Introducing the redefined 2-cocycle that depends on $c$ and $\kappa$,
\begin{align}
\tilde z_\bk(g,h):= (-1)^{\frac{1-c_g\kappa_g}{2} \frac{1-\kappa_h}{2}} z_\bk(g,h),\quad g,h \in G, 
\end{align}
the classification of point-gap topological phases can be written, in the notation of \cite{FreedTwisted2013}, as the degree $(-1)$ twisted equivariant $K$-group ${}^\phi K_G^{(\tilde z,\kappa)-1}(X)$\footnote{In the notation of \eqref{eq:equ_K_def_KP}, $X$ is a $G$-space and a left action $\alpha$ on $X$ is implicitly fixed.}:
\begin{align}
K_{\rm P}(G,X,\alpha,\phi,c,\kappa,z) = {}^\phi K_G^{(\tilde z,\kappa)-1}(X).
\label{eq:equ_K_def_KP}
\end{align}
As in Sec.~\ref{sec:P_L_top_phase}, real- and imaginary-line-gap topological phases are defined by imposing additional chiral symmetries $\s_y$ and $\s_x$, respectively.  
Their algebra with $U'_\bk(g)$ reads
\begin{align}
&U'_\bk(g) (i\s_y) = c_g (i\s_y) U'_\bk(g) ,\quad g \in G, \\
&U'_\bk(g) \s_x = c_g \kappa_g \s_x U'_\bk(g) ,\quad g \in G,
\end{align}
so in general they cannot be written as ${}^\phi K_G^{(\tilde z,\kappa)-n}(X)$ for a degree $n$.

However, in the following two cases the result reduces to a $K$-group with the original 2-cocycle $z$ at an appropriate degree.

\paragraph{AZ class.}
If $\kappa=c$, i.e., $\kappa_g=c_g$ for all $g\in G$, then $\tilde z=z$, and the classification of point-gap phases agrees with ${}^\phi K_G^{(z,c)-1}(X)$.  
Since the additional chiral symmetry $i\s_y$ corresponds to the $+1$ degree shift, the classification of real-line-gap phases agrees with ${}^\phi K_G^{(z,c)}(X)$. 
Thus,
\begin{align}
&K_{\rm P}(G,X,\alpha,\phi,c,\kappa,z)\big|_{\kappa=c}
={}^\phi K_G^{(z,c)-1}(X), \label{eq:iso_AZ_P}\\
&K_{\rm L_r}(G,X,\alpha,\phi,c,\kappa,z)\big|_{\kappa=c}
={}^\phi K_G^{(z,c)}(X). \label{eq:iso_AZ_Lr}
\end{align}
On the other hand, the additional chiral symmetry $\s_x$ cannot, in general, be absorbed as a degree shift.  
Hence the imaginary-line-gap phases cannot be expressed purely as a $K$-group for the symmetry group $G$ itself; rather, they are described by some $K$-group for the enlarged symmetry group $G\times \Z_2$ that includes the chiral symmetry $\s_x$.  
We refer to this setting as the AZ class.

\paragraph{AZ$^\dag$ class.}
If $\kappa = c \phi$, i.e., $\kappa_g=c_g\phi_g$ for all $g\in G$, redefine the symmetry by
\begin{align}
U''_\bk(g) = u_\bk(g) \otimes (\s_x)^{\frac{1-\phi_g c_g}{2}} .
\end{align}
Then the symmetry \eqref{eq:sym_KP} is equivalently written as
\begin{align}
\begin{cases}
U''_\bk(g)\tilde H_\bk^{\phi_g} U''_\bk(g)^\dag= c_g \tilde H_{g\bk},\quad g \in G, \\
U''_{h\bk}(g)U''_{\bk}(h)^{\phi_g}=z_\bk(g,h)U''_\bk(gh), \quad g,h \in G,\\
(i\sigma_z) \tilde H_\bk (i\sigma_z)^\dag = -\tilde H_{\bk}, \\
U''_\bk(g) (i \s_z)^{\phi_g}= c_g (i\s_z) U''_\bk(g),\quad g\in G.
\end{cases}
\end{align}
This is precisely the definition of the degree $+1$ shift via a chiral symmetry with $(i\s_z)^2=-1$.  
Moreover, the algebra
\begin{align}
U''_\bk(g) \s_y = c_g \s_y U''_\bk(g),\quad g \in G, 
\end{align}
shows that $\s_y$ acts as a chiral symmetry that lowers the degree by $1$.  
Therefore,
\begin{align}
&K_{\rm P}(G,X,\alpha,\phi,c,\kappa,z)\big|_{\kappa=\phi c}
={}^\phi K_G^{(z,c)+1}(X), \label{eq:iso_AZd_P}\\
&K_{\rm L_r}(G,X,\alpha,\phi,c,\kappa,z)\big|_{\kappa=\phi c}
={}^\phi K_G^{(z,c)}(X). \label{eq:iso_AZd_Lr}
\end{align}
As before, the additional chiral symmetry $\s_x$ cannot be absorbed as a degree shift, so in general the imaginary-line-gap phases cannot be written purely as a $K$-group for $G$; they are described instead by a suitable $K$-group for the enlarged symmetry group $G\times \Z_2$ including $\s_x$.  
We refer to this setting as the AZ$^\dag$ class.

\section{Classification of intrinsic topological phases with internal symmetries}
In this section, we present explicit computations of the homomorphisms $f_{\rm r}, f_{\rm i}$ for all possible internal symmetry classes, and provide the classification table of intrinsic topological phases.

\subsection{54 internal symmetry classes}
We first derive the set of independent internal symmetry classes.  
Since we are concerned only with internal symmetries, we may take the Hamiltonian $H$ to be a non-Hermitian matrix without parameter dependence.  
The symmetry conditions, for a finite group $G$, are given by
\begin{align}
\begin{cases}
u(g)\, H^{\phi_g,\kappa_g}\, u(g)^\dag = c_g H, \quad g \in G, \\
u(g)\, u(h)^{\phi_g} = z(g,h)\, u(gh), \quad g,h \in G,
\end{cases}
\end{align}
where $u(g)$ are unitary matrices and $z(g,h)$ is a 2-cocycle $z \in Z^2(G,U(1)_\phi)$.  
The matrix $H$ decomposes into blocks according to the irreducible representations of the unitary subgroup $G_0 = \{g \in G \mid \phi_g = \eta_g = c_g = 1\}$. 
For non-unitary elements $g \in G \setminus G_0$, the symmetry either permutes irreducible representations or acts effectively as a $\Z_2$ symmetry within a given block.

Our interest is in counting the number of inequivalent symmetry classes that close within an irreducible representation.  
Hence, without loss of generality, we may assume the symmetry group to be of the form $G = \Z_2^{\times N}$.  
We can also assume the absence of any element with $(\phi_g,\eta_g,c_g) = (1,1,1)$, so it suffices to consider $N=0,1,2,3$.  
We denote the symmetry matrices by $u_1,u_2,u_3$, etc., and analyze the inequivalent cases as follows.

\begin{itemize}
\item
$N=0$: There is only one trivial class.

\item
$N=1$: There are seven basic types as shown in Table~\ref{tab:BL_1}.
Since $\phi_g=-1$ allows two possibilities $u_1 u_1^* = \pm 1$, we obtain $2\times 4 + 3 = 11$ classes in total.
\begin{table}[]
    \caption{Symmetry types for $N=1$ generators.}
    \label{tab:BL_1}
    \centering
    $$
\begin{array}{ccc}
(\phi_g,c_g,\kappa_g) & \text{type} \\
\hline\hline
(-1,1,1) & \text{TRS} \\
(-1,1,-1) & \text{TRS}^\dag \\
(-1,-1,-1) & \text{PHS} \\
(-1,-1,1) & \text{PHS}^\dag \\
(1,-1,1) & \text{SLS} \\
(1,1,-1) & \text{pH} \\
(1,-1,-1) & \text{CS} \\
\end{array}
$$
\end{table}

\item
$N=2$:  
If $\phi_g=-1$ is included, the generators may be fixed to $\phi_1=\phi_2=-1$, i.e., one of TRS, PHS, TRS$^\dag$, or PHS$^\dag$.  
This gives six case as shown in Table~\ref{tab:BL_2}. 
The relative phase can be fixed by $u_1 u_2^* = u_2 u_1^*$.  
For each of the six cases, the choices $u_1 u_1^*=\pm 1$ and $u_2 u_2^*=\pm 1$ yield $2^2$ possibilities.  
If $\phi_g=-1$ is absent, only one case remains as shown in Table~\ref{tab:BL_2}. 
Here the commutation relation $u_1 u_2 = \pm u_2 u_1$ introduces two choices.  
Thus, the total is $2^2 \times 6 + 2 = 26$ classes.

\begin{table}[]
    \caption{Symmetry generator pairs for $N=2$.}
    \label{tab:BL_2}
    \centering
    $$
\begin{array}{cccc}
(\phi_g,c_g,\kappa_g) & \text{type} & \text{equivalent type} \\
\hline\hline
(-1,1,1),(-1,-1,-1) & \text{TRS+PHS} \\
(-1,1,-1),(-1,-1,1) & \text{TRS$^\dag$+PHS$^\dag$} \\
(-1,1,1),(-1,1,-1) & \text{TRS+TRS$^\dag$} & \text{TRS+pH} \\
(-1,1,1),(-1,-1,1) & \text{TRS+PHS$^\dag$} & \text{TRS+SLS} \\
(-1,-1,-1),(-1,-1,1) & \text{PHS+TRS$^\dag$} & \text{PHS+pH} \\
(-1,-1,-1),(-1,1,-1) & \text{PHS+PHS$^\dag$} & \text{PHS+SLS} \\
\hline
(1,-1,1),(1,1,-1) & \text{SLS+pH} & \text{CS+SLS, CS+pH} \\
\end{array}
$$
\end{table}

\item
$N=3$:  
The generators can be fixed as $\phi_1=\phi_2=\phi_3=-1$.  
This yields one case as shown in Table~\ref{tab:BL_3}. 
Relabeling this as the equivalent TRS+PHS+SLS case with generators $T,C,S$, one can fix the relative phase $u_T u_C^* = u_C u_T^*$ and set $u_S^2=1$.  
The remaining ambiguity lies in the signs of $u_T u_T^*, u_C u_C^*$ and the commutation relations $u_T u_S^* = \pm u_S u_T$, $u_C u_S^* = \pm u_S u_C$, giving $2^4=16$ classes.

\begin{table}[]
    \caption{Symmetry generator triple for $N=3$.}
    \label{tab:BL_3}
    \centering
$$\begin{array}{cccc}
(\phi_g,c_g,\kappa_g) & \text{type} & \text{equivalent type} \\
\hline\hline
(-1,1,1),(-1,-1,1),(-1,1,-1) & \text{TRS+PHS$^\dag$+TRS$^\dag$} & \text{TRS+PHS+SLS, TRS+PHS+pH} \\
\end{array}$$
\end{table}

\end{itemize}

From the above discussion, we obtain $1+11+26+16=54$ symmetry classes in total.  
This classification shows that these $54$ classes are specified by either (i) the AZ classes with $\kappa=c$, (ii) the real AZ$^\dag$ classes with $\kappa=\phi c$, or (iii) one of the AZ classes with an additional SLS or pH symmetry.  
For clarity, we make the definitions explicit below (our conventions follow \cite{KawabataClassification2019}).

\begin{itemize}
\item
For the 10 AZ classes, we consider TRS and/or PHS, together with CS:
\begin{align}
&{\rm TRS}: \quad T H^* T^\dag = H, \quad TT^* = \pm 1, \label{eq:AZTRS}\\
&{\rm PHS}: \quad C H^\top C^\dag = -H, \quad CC^* = \pm 1, \label{eq:AZPHS}\\
&{\rm CS}: \quad \Gamma H^\dag \Gamma^\dag = -H, \quad \Gamma^2 = 1. \label{eq:AZCS}
\end{align}

\item
For the 8 real AZ$^\dag$ classes, we consider TRS$^\dag$ and/or PHS$^\dag$:
\begin{align}
&{\rm TRS}^\dag: \quad T H^\top T^\dag = H, \quad TT^* = \pm 1, \\
&{\rm PHS}^\dag: \quad C H^* C^\dag = -H, \quad CC^* = \pm 1.
\end{align}

\item
For AZ classes with additional SLS or pH, we take the 10 AZ classes defined by 
(\ref{eq:AZTRS}), (\ref{eq:AZPHS}), (\ref{eq:AZCS}), and add either SLS or pH:
\begin{align}
&{\rm SLS}: \quad S H S^\dag = -H, \quad S^2 = 1, \\
&{\rm pH}: \quad \eta H^\dag \eta^\dag = H, \quad \eta^2 = 1.
\end{align}
We further specify the commutation relations with the AZ symmetries by signs:
\begin{align}
&TS^*=\epsilon_T ST, \quad CS^*=\epsilon_C SC, \quad \Gamma S = \epsilon_\Gamma S \Gamma, \quad 
\epsilon_T,\epsilon_C,\epsilon_\Gamma \in \{\pm 1\}, \\
&T\eta^*=\epsilon'_T \eta T, \quad C\eta^*=\epsilon'_C \eta C, \quad \Gamma \eta = \epsilon'_\Gamma \eta \Gamma, \quad 
\epsilon'_T,\epsilon'_C,\epsilon'_\Gamma \in \{\pm 1\}.
\end{align}
Using these signs, we label the symmetry classes as follows:  
if only TRS is present, $S_{\epsilon_T}$ or $\eta_{\epsilon'_T}$;  
if only PHS is present, $S_{\epsilon_C}$ or $\eta_{\epsilon'_C}$;  
if only CS is present, $S_{\epsilon_\Gamma}$ or $\eta_{\epsilon'_\Gamma}$;  
if both TRS and PHS are present, $S_{\epsilon_T\epsilon_C}$ or $\eta_{\epsilon'_T\epsilon'_C}$.
\end{itemize}

It is convenient to introduce labels as follows:  
$s \in \{0,1\}$ for the complex AZ classes (period $2$),  
$s \in \{0,\dots,7\}$ for the real AZ and real AZ$^\dag$ classes (period $8$),  
and $t$ to distinguish the additional symmetries:  
$t \in \{0,1\}$ for complex AZ classes with SLS or pH (period $2$),  
and $t \in \{0,1,2,3\}$ for real AZ classes with SLS or pH (period $4$).  
See Table.~\ref{tab:54}.

\begin{table}[!]
\begin{center}
\caption{Fifty-four internal symmetry classes. A comma in the table indicates ``or.''}
\label{tab:54}
\begin{tabular}{ccccccc}
&AZ/AZ$^\dag$ class&$s$\\
\hline
\multirow{2}{*}{Complex AZ}&A&0\\
&AIII&1\\
\hline
&AI&0\\
&BDI&1\\
&D&2\\
\multirow{2}{*}{Real AZ}&DIII&3\\
&AII&4\\
&CII&5\\
&C&6\\
&CI&7\\
\hline
&AI$^\dag$&0\\
&BDI$^\dag$&1\\
&D$^\dag$&2\\
\multirow{2}{*}{Real AZ$^\dag$}&DIII$^\dag$&3\\
&AII$^\dag$&4\\
&CII$^\dag$&5\\
&C$^\dag$&6\\
&CI$^\dag$&7\\
\end{tabular}
$$
\begin{array}{cccc}
\mbox{AZ class}&s\backslash t&0&1\\
\hline
{\rm A}&0&\eta&S\\
{\rm AIII}&1&S_+,\eta_+&S_-,\eta_-\\
\end{array}
$$
$$
\begin{array}{cccccc}
\mbox{AZ class}&s\backslash t&0&1&2&3\\
\hline
{\rm AI}&0&\eta_+&S_-&\eta_-&S_+\\
{\rm BDI}&1&S_{++},\eta_{++}&S_{-+},\eta_{+-}&S_{--},\eta_{--}&S_{+-},\eta_{-+}\\
{\rm D}&2&\eta_+&S_+&\eta_-&S_-\\
{\rm DIII}&3&S_{--},\eta_{++}&S_{-+},\eta_{-+}&S_{++},\eta_{--}&S_{+-},\eta_{+-}\\
{\rm AII}&4&\eta_+&S_-&\eta_-&S_+\\
{\rm CII}&5&S_{++},\eta_{++}&S_{-+},\eta_{+-}&S_{--},\eta_{--}&S_{+-},\eta_{-+}\\
{\rm C}&6&\eta_+&S_+&\eta_-&S_-\\
{\rm CI}&7&S_{--},\eta_{++}&S_{-+},\eta_{-+}&S_{++},\eta_{--}&S_{+-},\eta_{+-}\\
\end{array}
$$
\end{center}
\end{table}

The redefinition of the Hamiltonian $H \mapsto iH$ permutes the set of the 54 symmetry classes.  
The resulting correspondences are summarized in Tables~\ref{tab:HiH_1} and \ref{tab:HiH_2}. 

\begin{table}[]
    \caption{Symmetry class of $iH$ when $H$ belongs to an AZ or AZ$^\dag$ class.}
    \label{tab:HiH_1}
    \centering
    $$
\begin{array}{cc}
H & iH \\
\hline\hline
{\rm A}&{\rm A}\\
{\rm AIII}&{\rm A+\eta}\\
{\rm AI}&{\rm D^\dag}\\
{\rm BDI}&{\rm D+\eta_+}\\
{\rm D}&{\rm D}\\
{\rm DIII}&{\rm D+\eta_-}\\
{\rm AII}&{\rm C^\dag}\\
{\rm CII}&{\rm C+\eta_+}\\
{\rm C}&{\rm C}\\
{\rm CI}&{\rm C+\eta_-}\\
\hline
{\rm AI^\dag}&{\rm AI^\dag}\\
{\rm BDI^\dag}&{\rm AI+\eta_+}\\
{\rm D^\dag}&{\rm AI}\\
{\rm DIII^\dag}&{\rm AI+\eta_-}\\
{\rm AII^\dag}&{\rm AII^\dag}\\
{\rm CII^\dag}&{\rm AII+\eta_+}\\
{\rm C^\dag}&{\rm AII}\\
{\rm CI^\dag}&{\rm AII+\eta_-}\\
\end{array}
$$
\end{table}

\begin{table}[]
    \caption{Symmetry class of $iH$ when $H$ belongs to an AZ with an additional symmetry.}
    \label{tab:HiH_2}
    \centering
    $$
\begin{array}{cccc}
s&\mbox{AZ}&t=0&t=1\\
\hline
0&{\rm A}&\eta&S\\
1&{\rm AIII}&S_+,\eta_+&S_-,\eta_-\\
\end{array}
\quad \xrightarrow{H \mapsto iH} \quad 
\begin{array}{cccc}
s&t=0&t=1\\
\hline
0&{\rm AIII}&{\rm A+S}\\
1&{\rm AIII+S_+,\eta_+}&{\rm AIII+S_-,\eta_-}\\
\end{array}
$$
\begin{align*}
&\begin{array}{cccccc}
s&\mbox{AZ}&t=0&t=1&t=2&t=3\\
\hline
0&{\rm AI}&\eta_+&S_-&\eta_-&S_+\\
1&{\rm BDI}&S_{++},\eta_{++}&S_{-+},\eta_{+-}&S_{--},\eta_{--}&S_{+-},\eta_{-+}\\
2&{\rm D}&\eta_+&S_+&\eta_-&S_-\\
3&{\rm DIII}&S_{--},\eta_{++}&S_{-+},\eta_{-+}&S_{++},\eta_{--}&S_{+-},\eta_{+-}\\
4&{\rm AII}&\eta_+&S_-&\eta_-&S_+\\
5&{\rm CII}&S_{++},\eta_{++}&S_{-+},\eta_{+-}&S_{--},\eta_{--}&S_{+-},\eta_{-+}\\
6&{\rm C}&\eta_+&S_+&\eta_-&S_-\\
7&{\rm CI}&S_{--},\eta_{++}&S_{-+},\eta_{-+}&S_{++},\eta_{--}&S_{+-},\eta_{+-}\\
\end{array}\\
&\quad \xrightarrow{H \mapsto iH} \quad 
\begin{array}{cccccc}
s&t=0&t=1&t=2&t=3\\
\hline
0&{\rm BDI^\dag}&{\rm AII+S_-}&{\rm DIII^\dag}&{\rm AI+S_+}\\
1&{\rm BDI+S_{++},\eta_{++}}&{\rm DIII+S_{-+},\eta_{-+}}&{\rm DIII+S_{--},\eta_{++}}&{\rm BDI+S_{+-},\eta_{-+}}\\
2&{\rm BDI}&{\rm D+S_+}&{\rm DIII}&{\rm D+S_-}\\
3&{\rm BDI+S_{--},\eta_{--}}&{\rm BDI+S_{-+},\eta_{+-}}&{\rm DIII+S_{++},\eta_{--}}&{\rm DIII+S_{+-},\eta_{+-}}\\
4&{\rm CII^\dag}&{\rm AI+S_-}&{\rm CI^\dag}&{\rm AII+S_+}\\
5&{\rm CII+S_{++},\eta_{++}}&{\rm CI+S_{-+},\eta_{-+}}&{\rm CI+S_{--},\eta_{++}}&{\rm CII+S_{+-},\eta_{-+}}\\
6&{\rm CII}&{\rm C+S_+}&{\rm CI}&{\rm C+S_-}\\
7&{\rm CII+S_{--},\eta_{--}}&{\rm CII+S_{-+},\eta_{+-}}&{\rm CI+S_{++},\eta_{--}}&{\rm CI+S_{+-},\eta_{+-}}\\
\end{array}
\end{align*}
\end{table}

Accounting for the equivalence under $H \mapsto iH$, one finds 38 independent non-Hermitian symmetry classes~\cite{BernardClassification2002,KawabataClassification2019}~\footnote{Ref.~\cite{BernardClassification2002} initially listed 43 classes, later corrected.}.

\subsection{Classification of finite-dimensional non-Hermitian Hamiltonians}
In the remainder of this section, we compute the homomorphisms $f_{\rm r}, f_{\rm i}$ for all 54 internal symmetry classes in the case of non-Hermitian Hamiltonians depending on $d$ momentum-space variables $\R^d$ (arising from the Fourier transform of real-space coordinates) and on a parameter-space sphere $S^D$ characterizing real-space defects of codimension $D-1$~\cite{TeoTopological2010}.  
Since the map $H \mapsto iH$ permutes the symmetry classes, it suffices to compute the homomorphisms $f_{\rm r}$ from real-line-gap to point-gap phases; the maps $f_{\rm i}$ for imaginary-line-gap phases follow from this permutation.

For the full parameter space $\R^d \times S^D$, we compactify it to obtain $S^{d+D}$.  
We denote the projective coordinates on $S^{d+D}$ by $\bk = (k_1,\dots,k_d)$ and $\br=(r_1,\dots,r_D)$, and write the Hamiltonian as $H_{\bk,\br}$.  
The action of the 54 internal symmetries on the parameter space, specified by $\phi,c,\kappa$, is defined as
\begin{align}
u_g \, (H_{\bk,\br})^{\phi_g,\kappa_g} \, u_g^\dag = c_g H_{\phi_g \bk,\br}. 
\end{align}
In other words, symmetry with $\phi_g=-1$ is antiunitary and flips momentum variables, while leaving the defect parameters $\br$ invariant.
For each of the 54 symmetry classes, we denote the classification groups for point-gap, real-line-gap, and imaginary-line-gap phases as shown in Table~\ref{tab:Kgroup}. 

\begin{table}[]
    \caption{Symbols of classification groups for point-gap, real-line-gap, and imaginary-line-gap phases. }
    \label{tab:Kgroup}
    \centering
$$
\begin{array}{cccccc}
& \text{Point gap} & \text{Real-line gap} & \text{Imaginary-line gap} \\
\hline
\text{Complex AZ} & K^{\C}_{\rm P}(s;d,D) & K^{\C}_{\rm L_r}(s;d,D) & K^{\C}_{\rm L_i}(s;d,D) \\
\text{Real AZ} & K^{\R}_{\rm P}(s;d,D) & K^{\R}_{\rm L_r}(s;d,D) & K^{\R}_{\rm L_i}(s;d,D) \\
\text{Real AZ$^\dag$} & K^{\R^\dag}_{\rm P}(s;d,D) & K^{\R^\dag}_{\rm L_r}(s;d,D) & K^{\R^\dag}_{\rm L_i}(s;d,D) \\
\text{Complex AZ + SLS or pH} & K^{\C+U}_{\rm P}(s,t;d,D) & K^{\C+U}_{\rm L_r}(s,t;d,D) & K^{\C+U}_{\rm L_i}(s,t;d,D) \\
\text{Real AZ + SLS or pH} & K^{\R+U}_{\rm P}(s,t;d,D) & K^{\R+U}_{\rm L_r}(s,t;d,D) & K^{\R+U}_{\rm L_i}(s,t;d,D) \\
\end{array}
$$
\end{table}

We first identify the classification groups for point-gap and real-line-gap phases with the $K$-groups for the ten AZ classes of Hermitian systems~\cite{TeoTopological2010}, and with the $K$-groups for AZ classes with an additional $\Z_2$ unitary symmetry~\cite{ShiozakiTopology2014}.  
We then rewrite them, via suspension isomorphism, in terms of equivalent zero-dimensional $K$-groups.

Following \cite{TeoTopological2010}, we denote the $K$-groups for complex and real AZ classes of Hermitian systems as~\footnote{In \cite{TeoTopological2010}, these are denoted $K_{\C}(s;D,d)$ and $K_{\R}(s;D,d)$.}
\begin{align}
K^{\C}(s;d,D), \quad K^{\R}(s;d,D).
\end{align}
Similarly, following \cite{ShiozakiTopology2014}, we denote the $K$-groups for Hermitian AZ classes with an additional $\Z_2$ unitary symmetry by~\footnote{In \cite{ShiozakiTopology2014}, these are denoted $K_{\R}^{U}(s,t;d,d_\parallel,D,D_\parallel)$.}
\begin{align}
K^{\C+U}(s,t;d,d_\parallel,D,D_\parallel), \quad K^{\R+U}(s,t;d,d_\parallel,D,D_\parallel).
\end{align}
Here $d_\parallel$ and $D_\parallel$ denote the number of momentum variables $\bk$ and defect parameters $\br$ that are flipped by the additional $\Z_2$ symmetry, respectively.  
For convenience, we introduce
\begin{align}
\delta = d - D.
\end{align}
We also denote the classifying spaces for complex and real AZ classes by $C_s$ and $R_s$, respectively, as summarized in Table~\ref{tab:AZ_classifying_space}.

\begin{table}[!]
\begin{center}
\caption{Classifying space.}
\label{tab:AZ_classifying_space}
$$
\begin{array}{ccccc}
{\rm AZ\ class}&{\rm Classifying\ space}&{\rm Explicit\ form}&\pi_0(C_s) {\rm\ or\ } \pi_0(R_s) \\
\hline
{\rm A}&C_0&\frac{U(m+n)}{U(m) \times U(n)} \times \Z & \Z\\
{\rm AIII}&C_0&U(n)& 0\\
{\rm AI}&R_0&\frac{O(m+n)}{O(m) \times O(n)} \times \Z & \Z\\
{\rm BDI}&R_1&O(n)& \Z_2\\
{\rm D}&R_2&O(2n)/U(n)& \Z_2\\
{\rm DIII}&R_3&U(2n)/Sp(n)& 0\\
{\rm AII}&R_4&\frac{Sp(m+n)}{Sp(m) \times Sp(n)} \times \Z & 2\Z\\
{\rm CII}&R_5&Sp(n)& 0\\
{\rm C}&R_6&Sp(n)/U(n)& 0\\
{\rm CI}&R_7&U(n)/O(n)& 0\\
\end{array}
$$
\end{center}
\end{table}

For the AZ classes, it follows from \eqref{eq:iso_AZ_P} and \eqref{eq:iso_AZ_Lr} that
\begin{align}
    &K^{\mathbb{F}}_{\rm P}(s;d,D) 
    = K^{\mathbb{F}}(s+1;d,D) 
    \cong K^{\mathbb{F}}(s+1-\delta;0,0) 
    =
    \begin{cases}
        \pi_0(C_{s+1-\delta}) & (\mathbb{F}=\C), \\
        \pi_0(R_{s+1-\delta}) & (\mathbb{F}=\R), \\
    \end{cases} \\
    &K^{\mathbb{F}}_{\rm L_r}(s;d,D) 
    = K^{\mathbb{F}}(s;d,D) 
    \cong K^{\mathbb{F}}(s-\delta;0,0)
    =
    \begin{cases}
        \pi_0(C_{s-\delta}) & (\mathbb{F}=\C), \\
        \pi_0(R_{s-\delta}) & (\mathbb{F}=\R), \\
    \end{cases}
\end{align}

For the real AZ$^\dag$ classes, it follows from \eqref{eq:iso_AZd_P} and \eqref{eq:iso_AZd_Lr} that
\begin{align}
    &K^{\R^\dag}_{\rm P}(s;d,D) 
    = K^{\R}(s-1;d,D) 
    \cong K^{\R}(s-1-\delta;0,0) 
    = \pi_0(R_{s-1-\delta}),\\
    &K^{\R^\dag}_{\rm L_r}(s;d,D) 
    = K^{\R}(s;d,D) 
    \cong K^{\R}(s-\delta;0,0) 
    = \pi_0(R_{s-\delta}).
\end{align}

For the AZ classes with an additional SLS or pH, the symmetry classes $(s,t)$ have already been fixed in Table~\ref{tab:54} so that in the real-line-gap case they reduce to Hermitian AZ classes with an additional $\Z_2$ symmetry.  
We have
\begin{align}
    &K^{\mathbb{F}+U}_{\rm L_r}(s,t;d,D) 
    = K^{\mathbb{F}+U}(s,t;d,0,D,0) 
    \cong K^{\mathbb{F}+U}(s-\delta,t;0,0,0,0).
\end{align}
For point-gap phases, note that in the suspension construction \eqref{eq:sym_KP_suspension}, both TRS and PHS satisfy $c_g \kappa_g = 1$, so the additional coordinate $\theta$ is not flipped and thus behaves as an $\br$-type parameter.  
In contrast, the additional symmetries SLS and pH satisfy $c_g \kappa_g = -1$, so they flip the $\theta$-direction.  
Hence, the additional symmetry acts as a reflection along the new $(D+1)$ direction.  
We therefore obtain
\begin{align}
    K^{\mathbb{F}+U}_{\rm P}(s,t;d,D) 
    = K^{\mathbb{F}+U}(s,t;d,d_\parallel=0,D+1,D_\parallel=1)
    \cong K^{\mathbb{F}+U}(s-1-\delta,t-1;0,0,0,0), 
\end{align}
and in zero dimension, the $K$-groups are given by
\begin{align}
    &K^{\C+U}(s,t;0,0,0,0) = 
    \begin{cases}
        \pi_0(C_s \times C_s) & (t=0), \\
        \pi_0(C_{s+1}) & (t=1), \\
    \end{cases} \\
    &K^{\R+U}(s,t;0,0,0,0) = 
    \begin{cases}
        \pi_0(R_s \times R_s) & (t=0), \\
        \pi_0(R_{s-1}) & (t=1), \\
        \pi_0(C_s) & (t=2), \\
        \pi_0(R_{s+1}) & (t=3). \\
    \end{cases}
\end{align}

These results reproduce the classification obtained in \cite{KawabataSymmetry2019,ZhouPeriodic2019}.

\subsection{Homomorphism from real-line-gap to point-gap phases}
From the previous subsection, we have seen that both point-gap and real-line-gap phases reduce to the classification of zero-dimensional Hermitian systems.  
Therefore, to determine the homomorphism $f_{\rm r}$ from real-line-gap to point-gap phases in arbitrary dimensions, it suffices to compute it in zero dimension.  
By suspension isomorphism, the structure of the homomorphism is preserved and thus extends to finite-dimensional systems.

It is only necessary to compute the cases where both the zero-dimensional $K$-groups for real-line-gap and point-gap phases are nontrivial. 
There are 18 such cases. 
The results are summarized in Table~\ref{tab:SFH}.

\begin{table}[]
\caption{Building blocks of symmetry-forgetting homomorphisms. 
For each block, the second row represents the corresponding homomorphism from the imaginary-line gap to the point gap obtained by the redefinition $H \mapsto iH$.}
\label{tab:SFH}
\centering
$$
\begin{array}{ccccccccccccc}
\mbox{Symmetry class} & \mbox{Gap} & \mbox{$f_{\rm r}$ and $f_{\rm i}$}\\
\hline \hline
{\rm A+\eta}&{\rm L_r\to P}&\Z\oplus\Z\to\Z\\ 
{\rm AIII}&{\rm L_i\to P}&(n,m)\mapsto n+m\\
\hline
{\rm AI}&{\rm L_r\to P}&\Z\to\Z_2\\ 
{\rm D^\dag}&{\rm L_i\to P}&n\mapsto n\\
\hline
{\rm D^\dag}&{\rm L_r\to P}&\Z_2\to\Z_2\\ 
{\rm AI}&{\rm L_i\to P}&n\mapsto 0\\
\hline
{\rm BDI}&{\rm L_r\to P}&\Z_2\to\Z_2\\ 
{\rm D+\eta_+}&{\rm L_i\to P}&n\mapsto n\\
\hline
{\rm D+\eta_+}&{\rm L_r\to P}&\Z_2\oplus\Z_2\to\Z_2\\ 
{\rm BDI}&{\rm L_i\to P}&(n,m)\mapsto n+m\\
\hline
{\rm D+\eta_-}&{\rm L_r\to P}&\Z\to 2\Z\\ 
{\rm DIII}&{\rm L_i\to P}&n\mapsto n\\
\hline
{\rm C+\eta_-}&{\rm L_r\to P}&\Z\to \Z\\ 
{\rm CI}&{\rm L_i\to P}&n\mapsto 2n\\
\hline
{\rm AI+\eta_+}&{\rm L_r\to P}&\Z\oplus\Z\to \Z\\ 
{\rm BDI^\dag}&{\rm L_i\to P}&(n,m)\mapsto n+m\\
\hline
{\rm AI+\eta_-}&{\rm L_r\to P}&\Z\to \Z_2\\ 
{\rm DIII^\dag}&{\rm L_i\to P}&n\mapsto n\\
\hline
{\rm AII+\eta_+}&{\rm L_r\to P}&2\Z\oplus 2\Z\to 2\Z\\ 
{\rm CII^\dag}&{\rm L_i\to P}&(n,m)\mapsto n+m\\
\hline
{\rm AIII+S_-,\eta_-}&{\rm L_r\to P}&\Z\to \Z\oplus\Z\\ 
{\rm AIII+S_-,\eta_-}&{\rm L_i\to P}&n\mapsto (n,n)\\
\hline
{\rm AI+S_{+}}&{\rm L_r\to P}&\Z_2\to\Z_2\oplus\Z_2\\ 
{\rm AI+S_+}&{\rm L_i\to P}&n\mapsto (n,n)\\
\hline
{\rm BDI+S_{++},\eta_{++}}&{\rm L_r\to P}&\Z_2\oplus\Z_2\to\Z_2\\ 
{\rm BDI+S_{++},\eta_{++}}&{\rm L_i\to P}&(n,m)\mapsto n+m\\
\hline
{\rm BDI+S_{-+},\eta_{+-}}&{\rm L_r\to P}&\Z\to\Z\\ 
{\rm DIII+S_{-+},\eta_{-+}}&{\rm L_i\to P}&n\mapsto n\\
\hline
{\rm BDI+S_{+-},\eta_{-+}}&{\rm L_r\to P}&\Z_2\to\Z_2\oplus\Z_2\\ 
{\rm BDI+S_{+-},\eta_{-+}}&{\rm L_i\to P}&n\mapsto (n,n)\\
\hline
{\rm DIII+S_{+-},\eta_{+-}}&{\rm L_r\to P}&2\Z\to 2\Z\oplus 2\Z\\ 
{\rm DIII+S_{+-},\eta_{+-}}&{\rm L_i\to P}&n\mapsto (n,n)\\
\hline
{\rm CII+S_{-+},\eta_{+-}}&{\rm L_r\to P}&2\Z\to\Z\\ 
{\rm CI+S_{-+},\eta_{-+}}&{\rm L_i\to P}&n\mapsto 2n\\
\hline
{\rm CI+S_{+-},\eta_{+-}}&{\rm L_r\to P}&\Z\to \Z\oplus \Z\\ 
{\rm CI+S_{+-},\eta_{+-}}&{\rm L_i\to P}&n\mapsto (n,n)\\
\hline
\end{array}
$$
\end{table}

As for the zero-dimensional computation of $f_{\rm r}$, we proceed by hand: for each symmetry class, we present representative symmetry operators and pairs of Hamiltonians that generate the relevant $K$-groups in the Hermitian and non-Hermitian settings, and then identify the map by comparing these generators.  
When a pH $\eta H^\dag \eta^\dag = H$ is present, note the followings:
(i) In the Hermitian setting, the Hamiltonian decomposes into the sectors with $\eta=\pm 1$.
(ii) In the non-Hermitian setting, the product $\eta H$ is Hermitian, and analyzing the symmetries of $\eta H$ one can systematically construct generator models.

\subsubsection{A+$\eta$: $\Z \oplus \Z \to \Z$}
In the Hermitian case, the Hamiltonian decomposes into the sectors $\eta=\pm 1$.  
The generators of $\Z \oplus \Z$ are
\begin{align}
&(1,0): \qquad [\eta=1, \ H_0=1,\ H_1=-1], \\
&(0,1): \qquad [\eta=-1, \ H_0=1,\ H_1=-1]. 
\end{align}
In the non-Hermitian case, since $(\eta H)^\dag=\eta H$, the generator of $\Z$ is
\begin{align}
[\eta H_0=1,\ \eta H_1=-1].
\end{align}
Comparing the two, we obtain $\Z \oplus \Z \to \Z,\ (n,m) \mapsto n-m$.

\subsubsection{AI: $\Z \to \Z_2$}
The generators in both Hermitian and non-Hermitian cases are
\begin{align}
[T=1,\ H_0=1,\ H_1=-1]. 
\end{align}
Thus the map $\Z \to \Z_2$ is surjective.

\subsubsection{D$^\dag$: $\Z_2 \to \Z_2$}
A Hermitian generator is
\begin{align}
[C=\s_0,\ H_0=\s_y,\ H_1=-\s_y]. 
\end{align}
On the other hand, in the non-Hermitian case the generator is
\begin{align}
[C=1,\ H_0=i,\ H_1=-i].
\end{align}
Since the Hermitian generator corresponds to twice the non-Hermitian one, the map $\Z_2 \to \Z_2$ is the zero map.

\subsubsection{BDI: $\Z_2 \to \Z_2$}
The generators in both Hermitian and non-Hermitian cases are
\begin{align}
[T=1,\ C=\s_x,\ H_0=\s_z,\ H_1=-\s_z]. 
\end{align}
Hence the map $\Z_2\to \Z_2$ is surjective.  
(Note that no $1\times 1$ model with a point gap exists.)

\subsubsection{D+$\eta_+$: $\Z_2 \oplus \Z_2 \to \Z_2$}
The symmetries are
\begin{align}
&\eta H^\dag=H \eta,\quad \eta^2=1, \\
&CH^T=-HC,\quad CC^*=1, \\
&C\eta^*=\eta C.
\end{align}
In the Hermitian case, the Hamiltonian decomposes into $\eta=\pm 1$ sectors, with generators
\begin{align}
&[\eta=1,\ C=1,\ H_0=\s_y,\ H_1=-\s_y], \\
&[\eta=-1,\ C=1,\ H_0=\s_y,\ H_1=-\s_y]. 
\end{align}
In the non-Hermitian case, noting that
\begin{align}
\eta C (\eta H)^T = - (\eta H)\, \eta C,\qquad 
\eta C (\eta C)^* = 1,
\end{align}
we see that $\eta H$ follows the PHS of class D.  
Thus the generator is
\begin{align}
[\eta C=1,\ \eta H_0=\s_y,\ H_1=-\s_y].
\end{align}
Comparing them, the map is $\Z_2 \oplus \Z_2 \to \Z_2,\ (n,m)\mapsto n+m$.

\subsubsection{D+$\eta_-$: $\Z \to 2\Z$}
The symmetries are
\begin{align}
&\eta H^\dag=H \eta,\quad \eta^2=1, \\
&CH^T=-HC,\quad CC^*=1, \\
&C\eta^*=-\eta C.
\end{align}
In the Hermitian case, $H$ splits into $\eta=\pm 1$ sectors, and $C$ exchanges the sectors.  
A generator is
\begin{align}
[\eta=\s_y,\ C=1,\ H_0=\s_y,\ H_1=-\s_y].
\end{align}
In the non-Hermitian case, since
\begin{align}
\eta C (\eta H)^T = (\eta H)\,\eta C,\qquad 
\eta C (\eta C)^* = -1,
\end{align}
the operator $\eta H$ follows TRS of class AII.  
A generator is
\begin{align}
[\eta C=\s_y,\ \eta H_0=\s_0,\ \eta H_1=-\s_0].
\end{align}
Comparing them, the map $\Z\to 2\Z$ is an isomorphism.

\subsubsection{C+$\eta_-$: $\Z \to 2\Z$}
The symmetries are
\begin{align}
&\eta H^\dag = H \eta, \quad \eta^2=1, \\
&CH^T=-HC, \quad CC^*=-1, \\
&C\eta^*=-\eta C.
\end{align}
In the Hermitian case, $H$ splits into the $\eta=\pm 1$ sectors, and $C$ exchanges the sectors.  
A generator is
\begin{align}
[\eta=\s_y,\ C=\s_y,\ H_0=\s_y,\ H_1=-\s_y].
\end{align}
In the non-Hermitian case,
\begin{align}
\eta C (\eta H)^T = (\eta H)\,\eta C, \qquad 
\eta C (\eta C)^* = 1,
\end{align}
so $\eta H$ obeys TRS of class AI.  
A generator is
\begin{align}
[\eta C=1,\ \eta H_0=1,\ \eta H_1=-1].
\end{align}
Comparing them, the map $\Z \to \Z$ is multiplication by $2$.

\subsubsection{AI+$\eta_+$: $\Z \oplus \Z \to \Z$}
The symmetries are
\begin{align}
&\eta H^\dag = H \eta, \quad \eta^2=1, \\
&TH^*=HT, \quad TT^*=1, \\
&T\eta^*=\eta T.
\end{align}
In the Hermitian case, $H$ splits into $\eta=\pm 1$ sectors, and $T$ remains in each sector.  
A generator is
\begin{align}
[\eta=\pm 1,\ T=1,\ H_0=1,\ H_1=-1].
\end{align}
In the non-Hermitian case,
\begin{align}
T (\eta H)^* = (\eta H) T,\qquad 
TT^*=1,
\end{align}
so $\eta H$ follows TRS of class AI.  
A generator is
\begin{align}
[T=1,\ \eta H_0=1,\ \eta H_1=-1].
\end{align}
Thus the map is $\Z \oplus \Z \to \Z,\ (n,m)\mapsto n-m$.

\subsubsection{AI+$\eta_-$: $\Z \to \Z_2$}
The symmetries are
\begin{align}
&\eta H^\dag=H \eta, \quad \eta^2=1, \\
&TH^*=HT, \quad TT^*=1, \\
&T\eta^*=-\eta T.
\end{align}
In the Hermitian case, $H$ splits into $\eta=\pm 1$ sectors, and $T$ exchanges the sectors.  
A generator is
\begin{align}
[T=1,\ \eta=\s_y,\ H_0=\s_0,\ H_1=-\s_0].
\end{align}
In the non-Hermitian case,
\begin{align}
T (\eta H)^* = -(\eta H) T,\qquad 
TT^*=1,
\end{align}
so $\eta H$ follows PHS of class D.  
A generator is
\begin{align}
[T=1,\ \eta H_0=\s_y,\ \eta H_1=-\s_y].
\end{align}
Thus the map $\Z \to \Z_2$ is surjective.

\subsubsection{AII+$\eta_+$: $2\Z \oplus 2\Z \to 2\Z$}
The symmetries are
\begin{align}
&\eta H^\dag=H \eta, \quad \eta^2=1, \\
&TH^*=HT, \quad TT^*=-1, \\
&T\eta^*=\eta T.
\end{align}
In the Hermitian case, $H$ splits into $\eta=\pm 1$ sectors, and $T$ acts within each sector.  
A generator is
\begin{align}
[\eta=\pm \s_0,\ T=\s_y,\ H_0=\s_0,\ H_1=-\s_0].
\end{align}
In the non-Hermitian case,
\begin{align}
T (\eta H)^* = (\eta H) T,\qquad 
TT^*=-1,
\end{align}
so $\eta H$ follows TRS of class AII.  
A generator is
\begin{align}
[T=\s_y,\ \eta H_0=\pm \s_0,\ \eta H_1=\mp \s_0].
\end{align}
Thus the map is $2\Z \oplus 2\Z \to 2\Z,\ (n,m)\mapsto n-m$.

\subsubsection{AIII+$S_-,\eta_-$: $\Z \to \Z \oplus \Z$}
The symmetries are
\begin{align}
&\eta H^\dag=H \eta,\quad \eta^2=1, \\
&\G H^\dag=H \G,\quad \G^\dag=\G,\ \G^2=1, \\
&\G \eta = -\eta \G.
\end{align}
In the Hermitian case, $H$ splits into $\eta=\pm 1$ sectors, and $\G$ exchanges them.  
A generator is
\begin{align}
[\eta=\s_z,\ \G=\s_x,\ H_0=\s_z,\ H_1=-\s_z].
\end{align}
In the non-Hermitian case,
\begin{align}
\eta \G (\eta H) = (\eta H)\,\eta \G,\qquad 
(\eta \G)^2=-1,
\end{align}
so $\eta H$ decomposes into $\eta \G=\pm i$ sectors.  
A generator is
\begin{align}
[\eta \G=\pm i,\ \eta H_0=1,\ \eta H_1=-1].
\end{align}
Thus the map is $\Z \to \Z \oplus \Z,\ n \mapsto (n,n)$.

\subsubsection{AI+$S_+$: $\Z_2 \to \Z_2 \oplus \Z_2$}
The symmetries are
\begin{align}
&S H=-H S,\quad S^2=1, \\
&TH^*=HT,\quad TT^*=1, \\
&TS^*=ST.
\end{align}
In the Hermitian case, this is equivalent to class BDI, with generator
\begin{align}
[T=1,\ S=\s_z,\ H_0=\s_x,\ H_1=-\s_x].
\end{align}
In the non-Hermitian case, in the basis with $S=\s_z$ the Hamiltonian has the block form
\begin{align}
H=\begin{pmatrix}
&h_1\\
h_2\\
\end{pmatrix},\quad 
Th_j^*=h_j T,\quad j=0,1,
\end{align}
so a $\Z_2$ invariant ${\rm sign}[\det h_j]\in\{\pm 1\}$ is defined.  
Generators are
\begin{align}
[T=1,\ S=\s_z,\ H_0=\begin{pmatrix} &1\\ 1 \end{pmatrix},\ H_1=\begin{pmatrix} &-1\\ 1 \end{pmatrix}], \\
[T=1,\ S=\s_z,\ H_0=\begin{pmatrix} &1\\ 1 \end{pmatrix},\ H_1=\begin{pmatrix} &1\\ -1 \end{pmatrix}].
\end{align}
From the $\Z_2$ invariants of the Hermitian generator, the map is determined as $\Z_2 \to \Z_2 \oplus \Z_2,\ n \mapsto (n,n)$.

\subsubsection{BDI+$S_{++}, \eta_{++}$: $\Z_2 \oplus \Z_2 \to \Z_2$}
The symmetries are
\begin{align}
&\eta H^\dag = H \eta,\quad \eta^2=1, \\
&T H^* = H T,\quad TT^*=1, \\
&C H^T = -HC,\quad CC^*=1, \\
&T\eta^*=\eta T,\quad C \eta^*=\eta C,\quad TC^*=CT^*.
\end{align}
In the Hermitian case, $H$ splits into $\eta=\pm 1$ sectors, where $T$ and $C$ act within each sector as BDI symmetries.  
A generator is
\begin{align}
[\eta=\pm 1,\ T=1,\ C=\s_x,\ H_0=\s_z,\ H_1=-\s_z].
\end{align}
In the non-Hermitian case,
\begin{align}
&T (\eta H)^* = (\eta H) T,\qquad TT^*=1, \\
&\eta C (\eta H)^T = -(\eta H)\,\eta C,\qquad (\eta C)(\eta C)^* = 1,
\end{align}
so $\eta H$ belongs to class BDI.  
A generator is
\begin{align}
[T=1,\ C=\s_x,\ \eta H_0=\s_z,\ \eta H_1=-\s_z].
\end{align}
Comparing them, the map is $\Z_2 \oplus \Z_2 \to \Z_2,\ (n,m)\mapsto n+m$.

\subsubsection{BDI+$S_{-+}, \eta_{+-}$: $\Z \to \Z$}
The symmetries are
\begin{align}
&\eta H^\dag=H \eta,\quad \eta^2=1, \\
&T H^*=H T,\quad TT^*=1, \\
&C H^T=-HC,\quad CC^*=1, \\
&T\eta^*=\eta T,\quad C \eta^*=-\eta C,\quad TC^*=CT^*.
\end{align}
In the Hermitian case, $C$ exchanges the $\eta=\pm 1$ sectors, while $T$ remains in each sector, giving class AI.  
A generator is
\begin{align}
[\eta=\s_z,\ T=1,\ C=\s_x,\ H_0=\s_z,\ H_1=-\s_z].
\end{align}
In the non-Hermitian case, for $\eta H$ there is an additional $\Z_2$ symmetry
\begin{align}
\eta C T^\dag (\eta H) = (\eta H)\,\eta C T^\dag,\qquad (\eta C T^\dag)^2=-1,
\end{align}
together with TRS
\begin{align}
T (\eta H)^* = (\eta H) T,\qquad TT^*=1.
\end{align}
Since $T(\eta C T^\dag)^* = (\eta C T^\dag)T$, the operator $T$ exchanges the $\eta C T^\dag=\pm i$ sectors.  
A generator is
\begin{align}
[\eta C T^\dag=i\s_y,\ T=1,\ \eta H_0=\s_0,\ \eta H_1=-\s_0].
\end{align}
Comparing them, the map $\Z \to \Z$ is an isomorphism.

\subsubsection{BDI+$S_{+-}, \eta_{-+}$: $\Z_2 \to \Z_2 \oplus \Z_2$}
The symmetries are
\begin{align}
&\eta H^\dag=H \eta,\quad \eta^2=1, \\
&T H^*=H T,\quad TT^*=1, \\
&C H^T=-HC,\quad CC^*=1, \\
&T\eta^*=-\eta T,\quad C \eta^*=\eta C,\quad TC^*=CT^*.
\end{align}
In the Hermitian case, $T$ exchanges the $\eta=\pm 1$ sectors, and within each sector $C$ remains, giving class D.  
A generator is
\begin{align}
[\eta=\s_z \tau_0,\ T=\s_x \tau_0,\ C=\s_0 \tau_0,\ H_0=\s_0\tau_y,\ H_1=-\s_0\tau_y].
\end{align}
In the non-Hermitian case,
\begin{align}
&T (\eta H)^* = -\eta H T,\qquad TT^*=1, \\
&\eta C T^\dag (\eta H) = (\eta H)\,\eta C T^\dag,\qquad (\eta C T^\dag)^2=-1, \\
&T (\eta C T^\dag)^* = (\eta C T^\dag) T,
\end{align}
so $T$ acts within each $\eta C T^\dag=\pm i$ sector as a class D PHS.  
A generator is
\begin{align}
[T=1,\ \eta C T^\dag=\pm i,\ \eta H_0=\s_y,\ \eta H_1=-\s_y].
\end{align}
Comparing them, the map is $\Z_2 \to \Z_2 \oplus \Z_2,\ n \mapsto (n,n)$.

\subsubsection{DIII+$S_{+-}, \eta_{+-}$: $2\Z \to 2\Z \oplus 2\Z$}
The symmetries are
\begin{align}
&\eta H^\dag = H \eta,\quad \eta^2=1, \\
&TH^*=HT,\quad TT^*=-1, \\
&CH^\top=-HC,\quad CC^*=1, \\
&T\eta^*=\eta T,\quad C\eta^*=-\eta C,\quad TC^*=CT^*.
\end{align}
In the Hermitian case, $C$ exchanges the sectors $\eta=\pm 1$, while $T$ remains in each sector, giving class AII.  
A generator is
\begin{align}
[\eta=\s_0\tau_z,\ T=i\s_y\tau_0,\ C=\s_0\tau_x,\ H_0=\s_0\tau_z,\ H_1=-\s_0\tau_z].
\end{align}
In the non-Hermitian case,
\begin{align}
&T (\eta H)^* = \eta H\, T,\quad TT^*=-1, \\
&\eta C T^\dag\, (\eta H) = (\eta H)\, \eta C T^\dag,\quad (\eta C T^\dag)^2 = 1, \\
&T (\eta C T^\dag)^* = \eta C T^\dag\, T .
\end{align}
Hence $T$ acts within each $\Z_2$ sector labeled by $\eta C T^\dag=\pm 1$, and each sector is class AII.  
A generator is
\begin{align}
[T=\s_y,\ \eta C T^\dag=\pm 1,\ \eta H_0=\s_0,\ \eta H_1=-\s_0].
\end{align}
Comparing them, the map is $2\Z \to 2\Z \oplus 2\Z,\ n \mapsto (n,n)$.

\subsubsection{CII+$S_{-+}, \eta_{+-}$: $2\Z \to \Z$}
The symmetries are
\begin{align}
&\eta H^\dag = H \eta,\quad \eta^2=1, \\
&TH^*=HT,\quad TT^*=-1, \\
&CH^\top=-HC,\quad CC^*=-1, \\
&T\eta^*=\eta T,\quad C\eta^*=-\eta C,\quad TC^*=CT^*.
\end{align}
In the Hermitian case, $C$ exchanges the sectors $\eta=\pm 1$, while $T$ remains in each sector, giving class AII.  
A generator is
\begin{align}
[\eta=\s_0\tau_z,\ T=i\s_y\tau_0,\ C=i\s_0\tau_y,\ H_0=\s_0\tau_z,\ H_1=-\s_0\tau_z].
\end{align}
In the non-Hermitian case,
\begin{align}
&T (\eta H)^* = \eta H\, T,\quad TT^*=-1, \\
&\eta C T^\dag\, (\eta H) = (\eta H)\, \eta C T^\dag,\quad (\eta C T^\dag)^2 = -1, \\
&T (\eta C T^\dag)^* = \eta C T^\dag\, T .
\end{align}
Thus $T$ exchanges the sectors $\eta C T^\dag=\pm i$.  
A generator is
\begin{align}
[T=i\s_y,\ \eta C T^\dag=i\s_y,\ \eta H_0=\s_0,\ \eta H_1=-\s_0].
\end{align}
Comparing them, the map $2\Z \to \Z$ is $n \mapsto 2n$.

\subsubsection{CI+$S_{-+}, \eta_{+-}$: $\Z \to \Z \oplus \Z$}
The symmetries are
\begin{align}
&\eta H^\dag = H \eta,\quad \eta^2=1, \\
&TH^*=HT,\quad TT^*=1, \\
&CH^\top=-HC,\quad CC^*=-1, \\
&T\eta^*=\eta T,\quad C\eta^*=-\eta C,\quad TC^*=CT^*.
\end{align}
In the Hermitian case, $C$ exchanges the sectors $\eta=\pm 1$, while $T$ remains in each sector, giving class AI.  
A generator is
\begin{align}
[\eta=\s_z,\ T=\s_0,\ C=i\s_y,\ H_0=\s_z,\ H_1=-\s_z].
\end{align}
In the non-Hermitian case,
\begin{align}
&T (\eta H)^* = \eta H\, T,\quad TT^*=1, \\
&\eta C T^\dag\, (\eta H) = (\eta H)\, \eta C T^\dag,\quad (\eta C T^\dag)^2 = 1, \\
&T (\eta C T^\dag)^* = \eta C T^\dag\, T .
\end{align}
Therefore $T$ preserves each sector with $\eta C T^\dag=\pm 1$, giving class AI in each.  
A generator is
\begin{align}
[T=1,\ \eta C T^\dag=\pm 1,\ \eta H_0=1,\ \eta H_1=-1].
\end{align}
Comparing them, the map is $\Z \to \Z \oplus \Z,\ n \mapsto (n,n)$.

\subsection{When both $f_{\rm r}$ and $f_{\rm i}$ are $\Z \to \Z\oplus\Z$}
\label{sec:ZtoZZ}
As shown in the previous section, the homomorphisms from line-gap to point-gap phases $f_{\rm r}, f_{\rm i}$ can be independently obtained via suspension from the $0$-dimensional results.  
However, their \emph{relative relation} is not fixed in general.  
This ambiguity occurs when both $f_{\rm r}, f_{\rm i}$ take the form $\Z \to \Z\oplus\Z,\ n \mapsto (n,n)$.  
Once a basis of $\Z \oplus \Z$ is fixed, the maps $n \mapsto (n,n)$ and $n \mapsto (n,-n)$ are distinct homomorphisms.  
Thus, in such cases, the relative relation between $f_{\rm r}$ and $f_{\rm i}$ must be determined separately in each spatial dimension.  

The relevant cases are:  
(i) odd dimensions for symmetry classes containing A+S as a subgroup, and  
(ii) even dimensions for symmetry classes containing AIII+S$_-$ as a subgroup.  

\subsubsection{Odd dimensions: A+$S$}
For the symmetry class A+$S$, the Hamiltonian $H(\bk)$ can be written as
\begin{align}
H_\bk = 
\begin{pmatrix}
0 & h_1(\bk) \\
h_2(\bk) & 0
\end{pmatrix}_\s, 
\qquad 
S=\s_z,
\end{align}
where $h_1(\bk),h_2(\bk)$ are independent.  
In odd spatial dimensions $d=2n+1$, the winding numbers
\begin{align}
W^{(2n+1)}_i = \frac{n!}{(2 \pi i)^{n+1}(2n+1)!} \int \tr \bigl[h_i^{-1} d h_i \bigr]^{2n+1}, \quad i=1,2,
\label{eq:winding_num}
\end{align}
characterize the $\Z \oplus \Z$ classification of point-gap phases.  
With the real line gap, one may assume $H_\bk$ Hermitian without closing the gap, so that $h_2(\bk)^\dag = h_1(\bk)$.  
With the imaginary line gap, one may assume $H_\bk$ anti-Hermitian, so that $h_2(\bk)^\dag = -h_1(\bk)$.  
In both cases, the winding numbers satisfy $W^{(2n+1)}_2=-W^{(2n+1)}_1$.  
Hence both homomorphisms coincide: $f_{\rm r}=f_{\rm i}$.  
This can also be understood from the fact that class A+$S$ does not involve complex-conjugation symmetries.  
In the complex energy plane, the real and imaginary axes are not special lines; they are related by a continuous $90^\circ$ rotation $H_\bk \mapsto i H_\bk$.  

With an additional TRS commuting with $S$ (AI+S$_+$ and AII+S$_+$), the TRS $(T\otimes \s_0)K$ acts on each off-diagonal block $h_i(\bk)$ as $T h_i(\bk)^* = h_i(-\bk) T, TT^*=\pm 1$.  
For AI+S$_+$ and AII+S$_+$, the point-gap $K$-groups in $d=1,5$ are $\Z\oplus\Z$, characterized by the winding numbers \eqref{eq:winding_num}, and thus $f_{\rm r}=f_{\rm i}$ as in A+S.  

With an additional PHS anticommuting with $S$ (D+S$_-$ and C+S$_-$), the PHS $(C\otimes\s_y)$ acts on each off-diagonal block $h_i(\bk)$ as $C h_i(\bk)^T=-h_i(-\bk) C, CC^*=\pm 1$.  
For D+S$_-$ and C+S$_-$, the point-gap $K$-groups in $d=3,7$ are $\Z\oplus\Z$, characterized again by \eqref{eq:winding_num}, and thus $f_{\rm r}=f_{\rm i}$.  

\subsubsection{Even dimensions: AIII+$S_-$}
\label{sec:AIII+S_-}
For the symmetry class AIII+$S_-$, the Hamiltonian $H(\bk)$ is given by two Hermitian blocks $h_i(\bk)$ $(i=1,2)$:
\begin{align}
H(\bk) =
\begin{pmatrix}
0 & h_1(\bk) \\
h_2(\bk) & 0
\end{pmatrix}_\s, 
\qquad 
h_i(\bk)^\dag=h_i(\bk),\ i=1,2,
\label{eq:ham_aiii+S_-}
\end{align}
with $S=\s_z,\ \G=\s_y$.  
In even spatial dimensions $d=2n$, the $\Z\oplus\Z$ point-gap classification is given by the Chern numbers
\begin{align}
C^{(2n)}_i = \left(\frac{i}{2\pi}\right)^n \frac{1}{n!} \int \tr F_i^n,\quad i=1,2,
\label{eq:chern_num}
\end{align}
where $F_i$ is the Berry curvature for occupied states of $h_i(\bk)$.  
With the real line gap, $H(\bk)$ can be Hermitian, implying $h_2(\bk)=h_1(\bk)$ and hence $C^{(2n)}_2=C^{(2n)}_1$.  
Thus $f_{\rm r}: \Z \to \Z\oplus\Z$ is given by $1 \mapsto (1,1)$.  
With the imaginary line gap, $H(\bk)$ can be anti-Hermitian, implying $h_2(\bk)=-h_1(\bk)$ and hence $C^{(2n)}_2=-C^{(2n)}_1$.  
Thus $f_{\rm i}: \Z \to \Z\oplus\Z$ is given by $1 \mapsto (1,-1)$.  
Therefore, $f_{\rm r}$ and $f_{\rm i}$ differ in their relative relation.  

With an additional TRS commuting with both $S,\G$ (BDI+S$_{+-}$ and CII+S$_{+-}$), the TRS $(T\otimes\s_z)K$ acts on each block, imposing Hermitian PHS $T h_i(\bk)^*=-h_i(-\bk)T, TT^*=\pm 1$.  
In $d=2,6$, the point-gap groups are $\Z\oplus\Z$, characterized by \eqref{eq:chern_num}, and $f_{\rm r},f_{\rm i}$ are the same as for AIII+S$_-$.  

With an additional TRS commuting with $S$ but anticommuting with $\G$ (DIII+S$_{+-}$ and CI+S$_{+-}$), the TRS $(T\otimes\s_0)K$ acts on each block, imposing Hermitian TRS $T h_i(\bk)^*=h_i(-\bk)T, TT^*=\pm 1$.  
In $d=0,4$, the point-gap groups are again $\Z\oplus\Z$, characterized by \eqref{eq:chern_num}, and $f_{\rm r},f_{\rm i}$ behave as in AIII+S$_-$.

\subsection{Classification tables}

To obtain the homomorphisms from line-gap to point-gap phases, we first compute the maps 
$f_{\rm r}$ from real-line-gap phases and then add the equivalent symmetry classes related by 
the redefinition $H \mapsto iH$.  
The results for all 54 symmetry classes are summarized in Tables~\ref{tab:SFH_AZ} and 
\ref{tab:SFH_AZ_U}.  
Next, we reorganize the data from Tables~\ref{tab:SFH_AZ} and \ref{tab:SFH_AZ_U} to arrange 
the homomorphisms $f_{\rm r}, f_{\rm i}$ for each single symmetry class.  
The results are shown in Tables~\ref{tab:SFH_AZ_2} and \ref{tab:SFH_AZ_add_2}.
Note that when both $f_{\rm r},f_{\rm i}$ take the form $\Z \to \Z \oplus \Z$, the relative 
relation between them cannot be determined solely from the $0$-dimensional calculation as noted in Sec.~\ref{sec:ZtoZZ}.  
From these tables, we then obtain the classification tables for intrinsic non-Hermitian 
topological phases, given by the quotient groups $K_{\rm P}/(\im f_{\rm r} + \im f_{\rm i})$.  
The results are summarized in Tables~\ref{tab:SFH_quotient_AZ} and \ref{tab:SFH_quotient_AZ_add}.
Tables~\ref{tab:SFH_AZ_2},\ref{tab:SFH_AZ_add_2},\ref{tab:SFH_quotient_AZ} and \ref{tab:SFH_quotient_AZ_add} are the main results of this paper and were already announced in Supplemental Material in Ref.~\cite{OkumaTopological2020}.

\begin{sidewaystable}[htbp]
\caption{Real line gap to point gap homomorphisms for AZ and AZ$^\dag$ classes. 
Equivalent imaginary line gap to point gap homomorphisms by $H \to iH$ are shown in the row below each block.}
\label{tab:SFH_AZ}
\centering
{\small
$$
\begin{array}{ccccccccccccc}
\mbox{Symm. class} & \mbox{Gap} & \mbox{Class. space} &\delta=0&\delta=1&\delta=2&\delta=3&\delta=4&\delta=5&\delta=6&\delta=7 \\
\hline \hline
{\rm A}& {\rm L} \to {\rm P} & C_{-\delta} \to C_{1-\delta} &\Z \to 0&0 \to \Z&\Z \to 0&0 \to \Z&\Z \to 0&0 \to \Z&\Z \to 0&0 \to \Z\\ 
&&&&&&&&&&\\ 
\hline
{\rm AIII}& {\rm L_r}\to {\rm P}&C_{1-\delta}\to C_{-\delta}&0 \to \Z&\Z \to 0&0 \to \Z&\Z \to 0&0 \to \Z&\Z \to 0&0 \to \Z&\Z\to 0\\
{\rm A+\eta}& {\rm L_i} \to {\rm P}&&&&&&&&&\\ 
\hline
\hline 
{\rm AI}& {\rm L_r} \to {\rm P} & R_{-\delta} \to R_{1-\delta} &\Z \to \Z_2&0 \to \Z&0 \to 0&0 \to 0&2\Z\to 0&0\to 2\Z&\Z_2\to 0&\Z_2\to \Z_2 \\ 
{\rm D^\dag}& {\rm L_i} \to {\rm P} &&n \mapsto n&&&&&&&n\mapsto n \\ 
\hline
{\rm BDI}& {\rm L_r}\to {\rm P}&R_{1-\delta}\to R_{2-\delta}&\Z_2 \to \Z_2&\Z \to \Z_2&0 \to \Z&0 \to 0&0 \to 0&2\Z\to 0&0\to 2\Z&\Z_2\to 0\\ 
{\rm D+\eta_+}& {\rm L_i} \to {\rm P}&&n \mapsto n&n \mapsto n&&&&&&\\ 
\hline
{\rm D}& {\rm L} \to {\rm P}&R_{2-\delta}\to R_{3-\delta}&\Z_2 \to 0&\Z_2 \to \Z_2&\Z \to \Z_2&0 \to \Z&0 \to 0&0 \to 0&2\Z\to 0&0\to 2\Z\\ 
&&&&n \mapsto n&n \mapsto n&&&&&\\
\hline
{\rm DIII}& {\rm L_r}\to {\rm P}&R_{3-\delta}\to R_{4-\delta}&0\to 2\Z&\Z_2 \to 0&\Z_2 \to \Z_2&\Z \to \Z_2&0 \to \Z&0 \to 0&0 \to 0&2\Z\to 0\\ 
{\rm D+\eta_-}& {\rm L_i} \to {\rm P}&&&&n \mapsto n&n \mapsto n&&&&\\ 
\hline
{\rm AII}& {\rm L_r} \to {\rm P}&R_{4-\delta}\to R_{5-\delta}&2\Z \to 0&0\to 2\Z&\Z_2 \to 0&\Z_2 \to \Z_2&\Z \to \Z_2&0 \to \Z&0 \to 0&0 \to 0\\ 
{\rm C^\dag}& {\rm L_i} \to {\rm P}&&&&&n \mapsto n&n \mapsto n&&&\\ 
\hline
{\rm CII}& {\rm L_r} \to {\rm P}&R_{5-\delta}\to R_{6-\delta}&0 \to 0&2\Z \to 0&0\to 2\Z&\Z_2 \to 0&\Z_2 \to \Z_2&\Z \to \Z_2&0 \to \Z&0 \to 0\\ 
{\rm C+\eta_+}& {\rm L_i} \to {\rm P}&&&&&&n \mapsto n&n \mapsto n&&\\ 
\hline
{\rm C}& {\rm L} \to {\rm P}&R_{6-\delta}\to R_{7-\delta}&0 \to 0&0 \to 0&2\Z \to 0&0\to 2\Z&\Z_2 \to 0&\Z_2 \to \Z_2&\Z \to \Z_2&0 \to \Z\\ 
&&&&&&&&n \mapsto n&n \mapsto n&\\
\hline
{\rm CI}& {\rm L_r} \to {\rm P}&R_{7-\delta}\to R_{-\delta}&0 \to \Z&0 \to 0&0 \to 0&2\Z \to 0&0\to 2\Z&\Z_2 \to 0&\Z_2 \to \Z_2&\Z \to \Z_2\\ 
{\rm C+\eta_-}& {\rm L_i} \to {\rm P}&&&&&&&&n \mapsto n&n \mapsto n\\ 
\hline 
\hline
{\rm AI^\dag}& {\rm L} \to {\rm P}&R_{-\delta}\to R_{7-\delta}&\Z \to 0&0\to 0&0\to 0&0\to 2\Z&2\Z\to 0&0\to \Z_2&\Z_2\to \Z_2&\Z_2\to\Z\\ 
&&&&&&&&&n\mapsto 0&\\
\hline
{\rm BDI^\dag}&{\rm L_r} \to {\rm P}&R_{1-\delta}\to R_{-\delta}&\Z_2 \to \Z&\Z \to 0&0\to 0&0\to 0&0\to 2\Z&2\Z\to 0&0\to \Z_2&\Z_2\to \Z_2\\ 
{\rm AI+\eta_+}&{\rm L_i} \to {\rm P}&&&&&&&&&n\mapsto 0\\ 
\hline
{\rm D^\dag}&{\rm L_r} \to {\rm P}&R_{2-\delta}\to R_{1-\delta}&\Z_2 \to \Z_2&\Z_2 \to \Z&\Z \to 0&0\to 0&0\to 0&0\to 2\Z&2\Z\to 0&0\to \Z_2\\ 
{\rm AI}&{\rm L_i} \to {\rm P}&&n \mapsto 0&&&&&&&\\
\hline
{\rm DIII^\dag}& {\rm L_r} \to {\rm P}&R_{3-\delta}\to R_{2-\delta}&0 \to \Z_2&\Z_2 \to \Z_2&\Z_2 \to \Z&\Z \to 0&0\to 0&0\to 0&0\to 2\Z&2\Z\to 0\\ 
{\rm AI+\eta_-}& {\rm L_i} \to {\rm P}&&&n \mapsto 0&&&&&&\\
\hline
{\rm AII^\dag}& {\rm L} \to {\rm P}&R_{4-\delta}\to R_{3-\delta}&2\Z \to 0&0 \to \Z_2&\Z_2 \to \Z_2&\Z_2 \to \Z&\Z \to 0&0\to 0&0\to 0&0\to 2\Z\\
&&&&&n \mapsto 0&&&&&\\
\hline
{\rm CII^\dag}& {\rm L_r} \to {\rm P}&R_{5-\delta}\to R_{4-\delta}&0 \to 2\Z&2\Z \to 0&0 \to \Z_2&\Z_2 \to \Z_2&\Z_2 \to \Z&\Z \to 0&0\to 0&0\to 0\\ 
{\rm AII+\eta_+}&{\rm L_i} \to {\rm P}&&&&&n \mapsto 0&&&&\\
\hline
{\rm C^\dag}& {\rm L_r} \to {\rm P}&R_{6-\delta}\to R_{5-\delta}&0 \to 0&0 \to 2\Z&2\Z \to 0&0 \to \Z_2&\Z_2 \to \Z_2&\Z_2 \to \Z&\Z \to 0&0\to 0\\ 
{\rm AII}&{\rm L_i} \to {\rm P}&&&&&&n \mapsto 0&&&\\ 
\hline
{\rm CI^\dag}& {\rm L_r} \to {\rm P}&R_{7-\delta}\to R_{6-\delta}&0 \to 0&0 \to 0&0 \to 2\Z&2\Z \to 0&0 \to \Z_2&\Z_2 \to \Z_2&\Z_2 \to \Z&\Z \to 0\\ 
{\rm AII+\eta_-}&{\rm L_i} \to {\rm P}&&&&&&&n \mapsto 0&&\\
\hline
\end{array}
$$
}
\end{sidewaystable}

\begin{sidewaystable}[htbp]
\caption{Real line gap to point gap homomorphisms for AZ class with an additional symmetry.
Equivalent imaginary line gap to point gap homomorphisms by $H \to iH$ are shown in the row below each block.}
\label{tab:SFH_AZ_U}
\centering
{\tiny
$$
\begin{array}{ccccccccccccc}
\mbox{AZ}&\mbox{Add. symm.}&\mbox{Gap} & \mbox{Class. space} &\delta=0&\delta=1&\delta=2&\delta=3&\delta=4&\delta=5&\delta=6&\delta=7 \\
\hline \hline
{\rm A}&\eta&{\rm L_r} \to {\rm P}&C_{-\delta}\times C_{-\delta}\to C_{-\delta}&\Z\oplus\Z\to\Z&0\to 0\\ 
{\rm AIII}&&{\rm L_i} \to {\rm P}&&(n,m)\mapsto n-m&\\ 
\hline
{\rm AIII}&S_{+},\eta_{+}&{\rm L_r}\to {\rm P}&C_{1-\delta}\times C_{1-\delta} \to C_{1-\delta}&0\to 0&\Z\oplus\Z\to\Z\\ 
{\rm AIII}&S_+,\eta_+&{\rm L_i} \to {\rm P}&&&(n,m)\mapsto n-m\\ 
\hline \hline
{\rm A}&S&{\rm L} \to {\rm P}&C_{1-\delta}\to C_{1-\delta}\times C_{1-\delta}&0\to 0&\Z\to\Z\oplus\Z\\ 
&&&&&n\mapsto(n,n)\\ 
\hline
{\rm AIII}&S_-,\eta_-&{\rm L_r}\to {\rm P}&C_{-\delta}\to C_{-\delta}\times C_{-\delta}&\Z\to\Z\oplus\Z&0\to 0\\ 
{\rm AIII}&S_-,\eta_-&{\rm L_i} \to {\rm P}&&n\mapsto(n,n)&\\
\hline \hline
{\rm AI}&\rm \eta_+&{\rm L_r} \to {\rm P}&R_{-\delta}\times R_{-\delta} \to R_{-\delta}&\Z\oplus \Z\to\Z&0\to 0&0\to 0&0\to 0&2\Z\oplus 2\Z\to 2\Z&0\to 0&\Z_2\oplus\Z_2\to\Z_2&\Z_2\oplus\Z_2\to\Z_2\\ 
{\rm BDI^\dag}&&{\rm L_i} \to {\rm P}&&(n,m)\mapsto n+m&&&&(n,m)\mapsto n+m&&(n,m)\mapsto n+m&(n,m)\mapsto n+m\\ 
\hline
{\rm BDI}&S_{++},\eta_{++}&{\rm L_r}\to {\rm P}&R_{1-\delta}\times R_{1-\delta} \to R_{1-\delta}&\Z_2\oplus \Z_2\to\Z_2&\Z\oplus \Z\to\Z&0\to 0&0\to 0&0\to 0&2\Z\oplus 2\Z\to 2\Z&0\to 0&\Z_2\oplus\Z_2\to\Z_2\\ 
{\rm BDI}&S_{++},\eta_{++}&{\rm L_i} \to {\rm P}&&(n,m)\mapsto n+m&(n,m)\mapsto n+m&&&&(n,m)\mapsto n+m&&(n,m)\mapsto n+m\\ 
\hline
{\rm D}&\eta_+& {\rm L_r} \to {\rm P}&R_{2-\delta}\times R_{2-\delta} \to R_{2-\delta}&\Z_2\oplus\Z_2\to \Z_2&\Z_2\oplus \Z_2\to\Z_2&\Z\oplus \Z\to\Z&0\to 0&0\to 0&0\to 0&2\Z\oplus 2\Z\to 2\Z&0\to 0\\ 
{\rm BDI}&&{\rm L_i \to P}&&(n,m)\mapsto n+m&(n,m)\mapsto n+m&(n,m)\mapsto n+m&&&&(n,m)\mapsto n+m&\\
\hline
{\rm DIII}&S_{--},\eta_{++}& {\rm L_r}\to {\rm P}&R_{3-\delta}\times R_{3-\delta} \to R_{3-\delta}&0\to 0&\Z_2\oplus\Z_2\to \Z_2&\Z_2\oplus \Z_2\to\Z_2&\Z\oplus \Z\to\Z&0\to 0&0\to 0&0\to 0&2\Z\oplus 2\Z\to 2\Z\\ 
{\rm BDI}&S_{--},\eta_{--}& {\rm L_i} \to {\rm P}&&&(n,m)\mapsto n+m&(n,m)\mapsto n+m&(n,m)\mapsto n+m&&&&(n,m)\mapsto n+m\\ 
\hline
{\rm AII}&\eta_+& {\rm L_r} \to {\rm P}&R_{4-\delta}\times R_{4-\delta} \to R_{4-\delta}&2\Z\oplus 2\Z\to 2\Z&0\to 0&\Z_2\oplus\Z_2\to \Z_2&\Z_2\oplus \Z_2\to\Z_2&\Z\oplus \Z\to\Z&0\to 0&0\to 0&0\to 0\\ 
{\rm CII^\dag}&& {\rm L_i} \to {\rm P}&&(n,m)\mapsto n+m&&(n,m)\mapsto n+m&(n,m)\mapsto n+m&(n,m)\mapsto n+m&&\\
\hline
{\rm CII}&S_{++},\eta_{++}& {\rm L_r} \to {\rm P}&R_{5-\delta}\times R_{5-\delta} \to R_{5-\delta}&0\to 0&2\Z\oplus 2\Z\to 2\Z&0\to 0&\Z_2\oplus\Z_2\to \Z_2&\Z_2\oplus \Z_2\to\Z_2&\Z\oplus \Z\to\Z&0\to 0&0\to 0\\ 
{\rm CII}&S_{++},\eta_{++}& {\rm L_i} \to {\rm P}&&&(n,m)\mapsto n+m&&(n,m)\mapsto n+m&(n,m)\mapsto n+m&(n,m)\mapsto n+m&\\ 
\hline
{\rm C}&\eta_+& {\rm L_r} \to {\rm P}&R_{6-\delta}\times R_{6-\delta} \to R_{6-\delta}&0\to 0&0\to 0&2\Z\oplus 2\Z\to 2\Z&0\to 0&\Z_2\oplus\Z_2\to \Z_2&\Z_2\oplus \Z_2\to\Z_2&\Z\oplus \Z\to\Z&0\to 0\\ 
{\rm CII}&&{\rm L_i \to P}&&&&(n,m)\mapsto n+m&&(n,m)\mapsto n+m&(n,m)\mapsto n+m&(n,m)\mapsto n+m\\
\hline
{\rm CI}&S_{--},\eta_{++}& {\rm L_r} \to {\rm P}&R_{7-\delta}\times R_{7-\delta} \to R_{7-\delta}&0\to 0&0\to 0&0\to 0&2\Z\oplus 2\Z\to 2\Z&0\to 0&\Z_2\oplus\Z_2\to \Z_2&\Z_2\oplus \Z_2\to\Z_2&\Z\oplus \Z\to\Z\\ 
{\rm CII}&S_{--},\eta_{--}& {\rm L_i} \to {\rm P}&&&&&(n,m)\mapsto n+m&&(n,m)\mapsto n+m&(n,m)\mapsto n+m&(n,m)\mapsto n+m\\ 
\hline \hline
{\rm AI}&S_-&{\rm L_r \to \rm P}&R_{7-\delta}\to C_{1-\delta}&0\to 0&0\to\Z&0\to 0&2\Z\to\Z&0\to 0&\Z_2\to\Z&\Z_2\to 0&\Z\to\Z\\
{\rm AII}&S_-&{\rm L_i \to \rm P}&&&&&n\mapsto 2n&&&&n\mapsto n\\
\hline
{\rm BDI}&S_{-+},\eta_{+-}&{\rm L_r \to \rm P}&R_{-\delta}\to C_{-\delta}&\Z\to \Z&0\to 0&0\to\Z&0\to 0&2\Z\to\Z&0\to 0&\Z_2\to\Z&\Z_2\to 0\\
{\rm DIII}&S_{-+},\eta_{-+}&{\rm L_i \to \rm P}&&n\mapsto n&&&&n\mapsto 2n&&&\\
\hline
{\rm D}&S_+&{\rm L \to \rm P}&R_{1-\delta}\to C_{1-\delta}&\Z_2\to 0&\Z\to \Z&0\to 0&0\to\Z&0\to 0&2\Z\to\Z&0\to 0&\Z_2\to\Z\\
&&&&&n\mapsto n&&&&n\mapsto 2n&&\\
\hline
{\rm DIII}&S_{-+},\eta_{-+}&{\rm L_r \to \rm P}&R_{2-\delta}\to C_{-\delta}&\Z_2\to \Z&\Z_2\to 0&\Z\to \Z&0\to 0&0\to\Z&0\to 0&2\Z\to\Z&0\to 0\\
{\rm BDI}&S_{-+},\eta_{+-}&{\rm L_i \to \rm P}&&&&n\mapsto n&&&&n\mapsto 2n&\\
\hline
{\rm AII}&S_-&{\rm L_r \to \rm P}&R_{3-\delta}\to C_{1-\delta}&0\to 0&\Z_2\to \Z&\Z_2\to 0&\Z\to \Z&0\to 0&0\to\Z&0\to 0&2\Z\to\Z\\
{\rm AI}&S_-&{\rm L_i \to \rm P}&&&&&n\mapsto n&&&&n\mapsto 2n\\
\hline
{\rm CII}&S_{-+},\eta_{+-}&{\rm L_r \to \rm P}&R_{4-\delta}\to C_{-\delta}&2\Z\to \Z&0\to 0&\Z_2\to \Z&\Z_2\to 0&\Z\to \Z&0\to 0&0\to\Z&0\to 0\\
{\rm CI}&S_{-+},\eta_{-+}&{\rm L_i \to \rm P}&&n\mapsto 2n&&&&n\mapsto n&&&\\
\hline
{\rm C}&S_+&{\rm L \to \rm P}&R_{5-\delta}\to C_{1-\delta}&0\to 0&2\Z\to \Z&0\to 0&\Z_2\to \Z&\Z_2\to 0&\Z\to \Z&0\to 0&0\to\Z\\
&&&&&n\mapsto 2n&&&&n\mapsto n&&\\
\hline
{\rm CI}&S_{-+},\eta_{-+}&{\rm L_r \to \rm P}&R_{6-\delta}\to C_{-\delta}&0\to \Z&0\to 0&2\Z\to \Z&0\to 0&\Z_2\to \Z&\Z_2\to 0&\Z\to \Z&0\to 0\\
{\rm CII}&S_{-+},\eta_{+-}&{\rm L_i \to \rm P}&&&&n\mapsto 2n&&&&n\mapsto n&\\
\hline \hline
{\rm AI}&\rm \eta_-&{\rm L_r} \to {\rm P}&C_{-\delta}\to R_{2-\delta}&\Z\to\Z_2&0\to\Z_2&\Z\to\Z&0\to 0&\Z\to 0&0\to 0&\Z\to 2\Z&0\to 0\\ 
{\rm DIII^\dag}&&{\rm L_i} \to {\rm P}&&n\mapsto n&&n\mapsto 2n&&&&n\mapsto n&\\ 
\hline
{\rm BDI}&S_{--},\eta_{--}&{\rm L_r}\to {\rm P}&C_{1-\delta}\to R_{3-\delta}&0\to 0&\Z\to\Z_2&0\to\Z_2&\Z\to\Z&0\to 0&\Z\to 0&0\to 0&\Z\to 2\Z\\ 
{\rm DIII}&S_{--},\eta_{++}&{\rm L_i} \to {\rm P}&&&n\mapsto n&&n\mapsto 2n&&&&n\mapsto n\\ 
\hline
{\rm D}&\eta_-& {\rm L_r} \to {\rm P}&C_{-\delta}\to R_{4-\delta}&\Z\to 2\Z&0\to 0&\Z\to\Z_2&0\to\Z_2&\Z\to\Z&0\to 0&\Z\to 0&0\to 0\\ 
{\rm DIII}&&{\rm L_i \to P}&&n\mapsto n&&n\mapsto n&&n\mapsto 2n&&\\
\hline
{\rm DIII}&S_{++},\eta_{--}& {\rm L_r}\to {\rm P}&C_{1-\delta}\to R_{5-\delta}&0\to 0&\Z\to 2\Z&0\to 0&\Z\to\Z_2&0\to\Z_2&\Z\to\Z&0\to 0&\Z\to 0\\ 
{\rm DIII}&S_{++},\eta_{--}& {\rm L_i} \to {\rm P}&&&n\mapsto n&&n\mapsto n&&n\mapsto 2n&&\\ 
\hline
{\rm AII}&\eta_-& {\rm L_r} \to {\rm P}&C_{-\delta}\to R_{6-\delta}&\Z\to 0&0\to 0&\Z\to 2\Z&0\to 0&\Z\to\Z_2&0\to\Z_2&\Z\to\Z&0\to 0\\ 
{\rm CI^\dag}&& {\rm L_i} \to {\rm P}&&&&n\mapsto n&&n\mapsto n&&n\mapsto 2n&\\
\hline
{\rm CII}&S_{--},\eta_{--}& {\rm L_r} \to {\rm P}&C_{1-\delta}\to R_{7-\delta}&0\to 0&\Z\to 0&0\to 0&\Z\to 2\Z&0\to 0&\Z\to\Z_2&0\to\Z_2&\Z\to\Z\\ 
{\rm CI}&S_{--},\eta_{++}& {\rm L_i} \to {\rm P}&&&&&n\mapsto n&&n\mapsto n&&n\mapsto 2n\\ 
\hline
{\rm C}&\eta_-& {\rm L_r} \to {\rm P}&C_{-\delta}\to R_{-\delta}&\Z\to \Z&0\to 0&\Z\to 0&0\to 0&\Z\to 2\Z&0\to 0&\Z\to\Z_2&0\to\Z_2\\ 
{\rm CI}&&{\rm L_i \to P}&&n\mapsto 2n&&&&n\mapsto n&&n\mapsto n\\
\hline
{\rm CI}&S_{++},\eta_{--}& {\rm L_r} \to {\rm P}&C_{1-\delta}\to R_{1-\delta}&0\to \Z_2&\Z\to \Z&0\to 0&\Z\to 0&0\to 0&\Z\to 2\Z&0\to 0&\Z\to\Z_2\\ 
{\rm CI}&S_{++},\eta_{--}& {\rm L_i} \to {\rm P}&&&n\mapsto 2n&&&&n\mapsto n&&n\mapsto n\\ 
\hline \hline
{\rm AI}&S_+&{\rm L_r \to \rm P}&R_{1-\delta}\to R_{1-\delta}\times R_{1-\delta}&\Z_2\to\Z_2\oplus\Z_2&\Z\to\Z\oplus\Z&0\to 0&0\to 0&0\to 0&\Z\to\Z\oplus\Z&0\to 0&\Z_2\to\Z_2\oplus\Z_2\\
{\rm AI}&S_+&{\rm L_i \to \rm P}&&n\mapsto(n,n)&n\mapsto(n,n)&&&&n\mapsto(n,n)&&n\mapsto(n,n)\\
\hline
{\rm BDI}&S_{+-},\eta_{-+}&{\rm L_r \to \rm P}&R_{2-\delta}\to R_{2-\delta}\times R_{2-\delta}&\Z_2\to\Z_2\oplus\Z_2&\Z_2\to\Z_2\oplus\Z_2&\Z\to\Z\oplus\Z&0\to 0&0\to 0&0\to 0&\Z\to\Z\oplus\Z&0\to 0\\
{\rm BDI}&S_{+-},\eta_{-+}&{\rm L_i \to \rm P}&&n\mapsto(n,n)&n\mapsto(n,n)&n\mapsto(n,n)&&&&n\mapsto(n,n)&\\
\hline
{\rm D}&S_-&{\rm L \to \rm P}&R_{3-\delta}\to R_{3-\delta}\times R_{3-\delta}&0\to 0&\Z_2\to\Z_2\oplus\Z_2&\Z_2\to\Z_2\oplus\Z_2&\Z\to\Z\oplus\Z&0\to 0&0\to 0&0\to 0&\Z\to\Z\oplus\Z\\
&&&&&n\mapsto(n,n)&n\mapsto(n,n)&n\mapsto(n,n)&&&&n\mapsto(n,n)\\
\hline
{\rm DIII}&S_{+-},\eta_{+-}&{\rm L_r \to \rm P}&R_{4-\delta}\to R_{4-\delta}\times R_{4-\delta}&\Z\to\Z\oplus\Z&0\to 0&\Z_2\to\Z_2\oplus\Z_2&\Z_2\to\Z_2\oplus\Z_2&\Z\to\Z\oplus\Z&0\to 0&0\to 0&0\to 0\\
{\rm DIII}&S_{+-},\eta_{+-}&{\rm L_i \to \rm P}&&n\mapsto(n,n)&&n\mapsto(n,n)&n\mapsto(n,n)&n\mapsto(n,n)&&&\\
\hline
{\rm AII}&S_+&{\rm L_r \to \rm P}&R_{5-\delta}\to R_{5-\delta}\times R_{5-\delta}&0\to 0&\Z\to\Z\oplus\Z&0\to 0&\Z_2\to\Z_2\oplus\Z_2&\Z_2\to\Z_2\oplus\Z_2&\Z\to\Z\oplus\Z&0\to 0&0\to 0\\
{\rm AII}&S_+&{\rm L_i \to \rm P}&&&n\mapsto(n,n)&&n\mapsto(n,n)&n\mapsto(n,n)&n\mapsto(n,n)&&\\
\hline
{\rm CII}&S_{+-},\eta{-+}&{\rm L_r \to \rm P}&R_{6-\delta}\to R_{6-\delta}\times R_{6-\delta}&0\to 0&0\to 0&\Z\to\Z\oplus\Z&0\to 0&\Z_2\to\Z_2\oplus\Z_2&\Z_2\to\Z_2\oplus\Z_2&\Z\to\Z\oplus\Z&0\to 0\\
{\rm CII}&S_{+-},\eta_{-+}&{\rm L_i \to \rm P}&&&&n\mapsto(n,n)&&n\mapsto(n,n)&n\mapsto(n,n)&n\mapsto(n,n)&\\
\hline
{\rm C}&S_-&{\rm L \to \rm P}&R_{7-\delta}\to R_{7-\delta}\times R_{7-\delta}&0\to 0&0\to 0&0\to 0&\Z\to\Z\oplus\Z&0\to 0&\Z_2\to\Z_2\oplus\Z_2&\Z_2\to\Z_2\oplus\Z_2&\Z\to\Z\oplus\Z\\
&&&&&&&n\mapsto(n,n)&&n\mapsto(n,n)&n\mapsto(n,n)&n\mapsto(n,n)\\
\hline
{\rm CI}&S_{+-},\eta_{+-}&{\rm L_r \to \rm P}&R_{-\delta}\to R_{-\delta}\times R_{-\delta}&Z\to\Z\oplus\Z&0\to 0&0\to 0&0\to 0&\Z\to\Z\oplus\Z&0\to 0&\Z_2\to\Z_2\oplus\Z_2&\Z_2\to\Z_2\oplus\Z_2\\
{\rm CI}&S_{+-},\eta_{+-}&{\rm L_i \to \rm P}&&n\mapsto(n,n)&&&&n\mapsto(n,n)&&n\mapsto(n,n)&n\mapsto(n,n)\\
\hline \hline
\end{array}
$$
}
\end{sidewaystable}

\begin{sidewaystable}[!]
\caption{Line to point gap homomorphisms for AZ and AZ$^\dag$ classes.}
\label{tab:SFH_AZ_2}
\centering
{\scriptsize
$$
\begin{array}{ccccccccccccc}
\mbox{Symm. class} & \mbox{Gap} & \delta=0&\delta=1&\delta=2&\delta=3&\delta=4&\delta=5&\delta=6&\delta=7 \\
\hline \hline
{\rm A}& {\rm L} \to {\rm P} &\mathbb{Z} \to 0&0 \to \mathbb{Z}&\mathbb{Z} \to 0&0 \to \mathbb{Z}&\mathbb{Z} \to 0&0 \to \mathbb{Z}&\mathbb{Z} \to 0&0 \to \mathbb{Z}\\ 
&&&&&&&&&\\ 
\hline
{\rm AIII}& {\rm L_r}\to {\rm P}&0 \to \mathbb{Z}&\mathbb{Z} \to 0&0 \to \mathbb{Z}&\mathbb{Z} \to 0&0 \to \mathbb{Z}&\mathbb{Z} \to 0&0 \to \mathbb{Z}&\mathbb{Z} \to 0\\
&&&&&&&&&\\ 
&{\rm L_i} \to {\rm P}&\mathbb{Z}\oplus\mathbb{Z}\to\mathbb{Z}&0\to 0&\mathbb{Z}\oplus\mathbb{Z}\to\mathbb{Z}&0\to 0&\mathbb{Z}\oplus\mathbb{Z}\to\mathbb{Z}&0\to 0&\mathbb{Z}\oplus\mathbb{Z}\to\mathbb{Z}&0\to 0\\ 
&&(n,m)\mapsto n-m&&(n,m)\mapsto n-m&&(n,m)\mapsto n-m&&(n,m)\mapsto n-m&\\ 
\hline
{\rm AI}& {\rm L_r} \to {\rm P} &\mathbb{Z} \to \mathbb{Z}_2&0 \to \mathbb{Z}&0 \to 0&0 \to 0&2\mathbb{Z}\to 0&0\to 2\mathbb{Z}&\mathbb{Z}_2\to 0&\mathbb{Z}_2\to \mathbb{Z}_2 \\ 
&&n \mapsto n&&&&&&&n\mapsto n \\ 
&{\rm L_i} \to {\rm P}&\mathbb{Z}_2 \to \mathbb{Z}_2&\mathbb{Z}_2 \to \mathbb{Z}&\mathbb{Z} \to 0&0\to 0&0\to 0&0\to 2\mathbb{Z}&2\mathbb{Z}\to 0&0\to \mathbb{Z}_2\\ 
&&n \mapsto 0&&&&&&&\\
\hline
{\rm BDI}& {\rm L_r}\to {\rm P}&\mathbb{Z}_2 \to \mathbb{Z}_2&\mathbb{Z} \to \mathbb{Z}_2&0 \to \mathbb{Z}&0 \to 0&0 \to 0&2\mathbb{Z}\to 0&0\to 2\mathbb{Z}&\mathbb{Z}_2\to 0\\ 
&&n \mapsto n&n \mapsto n&&&&&&\\ 
&{\rm L_i \to P}&\mathbb{Z}_2\oplus\mathbb{Z}_2\to \mathbb{Z}_2&\mathbb{Z}_2\oplus \mathbb{Z}_2\to\mathbb{Z}_2&\mathbb{Z}\oplus \mathbb{Z}\to\mathbb{Z}&0\to 0&0\to 0&0\to 0&2\mathbb{Z}\oplus 2\mathbb{Z}\to 2\mathbb{Z}&0\to 0\\ 
&&(n,m)\mapsto n+m&(n,m)\mapsto n+m&(n,m)\mapsto n+m&&&&(n,m)\mapsto n+m&\\
\hline
{\rm D}& {\rm L} \to {\rm P}&\mathbb{Z}_2 \to 0&\mathbb{Z}_2 \to \mathbb{Z}_2&\mathbb{Z} \to \mathbb{Z}_2&0 \to \mathbb{Z}&0 \to 0&0 \to 0&2\mathbb{Z}\to 0&0\to 2\mathbb{Z}\\ 
&&&n \mapsto n&n \mapsto n&&&&&\\
\hline
{\rm DIII}& {\rm L_r}\to {\rm P}&0\to 2\mathbb{Z}&\mathbb{Z}_2 \to 0&\mathbb{Z}_2 \to \mathbb{Z}_2&\mathbb{Z} \to \mathbb{Z}_2&0 \to \mathbb{Z}&0 \to 0&0 \to 0&2\mathbb{Z}\to 0\\ 
&&&&n \mapsto n&n \mapsto n&&&&\\ 
&{\rm L_i \to P}&\mathbb{Z}\to 2\mathbb{Z}&0\to 0&\mathbb{Z}\to\mathbb{Z}_2&0\to\mathbb{Z}_2&\mathbb{Z}\to\mathbb{Z}&0\to 0&\mathbb{Z}\to 0&0\to 0\\ 
&&n\mapsto n&&n\mapsto n&&n\mapsto 2n&&\\
\hline
{\rm AII}& {\rm L_r} \to {\rm P}&2\mathbb{Z} \to 0&0\to 2\mathbb{Z}&\mathbb{Z}_2 \to 0&\mathbb{Z}_2 \to \mathbb{Z}_2&\mathbb{Z} \to \mathbb{Z}_2&0 \to \mathbb{Z}&0 \to 0&0 \to 0\\ 
&&&&&n \mapsto n&n \mapsto n&&&\\ 
&{\rm L_i} \to {\rm P}&0 \to 0&0 \to 2\mathbb{Z}&2\mathbb{Z} \to 0&0 \to \mathbb{Z}_2&\mathbb{Z}_2 \to \mathbb{Z}_2&\mathbb{Z}_2 \to \mathbb{Z}&\mathbb{Z} \to 0&0\to 0\\ 
&&&&&&n \mapsto 0&&&\\ 
\hline
{\rm CII}& {\rm L_r} \to {\rm P}&0 \to 0&2\mathbb{Z} \to 0&0\to 2\mathbb{Z}&\mathbb{Z}_2 \to 0&\mathbb{Z}_2 \to \mathbb{Z}_2&\mathbb{Z} \to \mathbb{Z}_2&0 \to \mathbb{Z}&0 \to 0\\ 
&&&&&&n \mapsto n&n \mapsto n&&\\ 
&{\rm L_i} \to {\rm P}&0\to 0&0\to 0&2\mathbb{Z}\oplus 2\mathbb{Z}\to 2\mathbb{Z}&0\to 0&\mathbb{Z}_2\oplus\mathbb{Z}_2\to \mathbb{Z}_2&\mathbb{Z}_2\oplus \mathbb{Z}_2\to\mathbb{Z}_2&\mathbb{Z}\oplus \mathbb{Z}\to\mathbb{Z}&0\to 0\\ 
&&&&(n,m)\mapsto n+m&&(n,m)\mapsto n+m&(n,m)\mapsto n+m&(n,m)\mapsto n+m\\
\hline
{\rm C}& {\rm L} \to {\rm P}&0 \to 0&0 \to 0&2\mathbb{Z} \to 0&0\to 2\mathbb{Z}&\mathbb{Z}_2 \to 0&\mathbb{Z}_2 \to \mathbb{Z}_2&\mathbb{Z} \to \mathbb{Z}_2&0 \to \mathbb{Z}\\ 
&&&&&&&n \mapsto n&n \mapsto n&\\
\hline
{\rm CI}& {\rm L_r} \to {\rm P}&0 \to \mathbb{Z}&0 \to 0&0 \to 0&2\mathbb{Z} \to 0&0\to 2\mathbb{Z}&\mathbb{Z}_2 \to 0&\mathbb{Z}_2 \to \mathbb{Z}_2&\mathbb{Z} \to \mathbb{Z}_2\\ 
&&&&&&&&n \mapsto n&n \mapsto n\\ 
& {\rm L_i} \to {\rm P}&\mathbb{Z}\to \mathbb{Z}&0\to 0&\mathbb{Z}\to 0&0\to 0&\mathbb{Z}\to 2\mathbb{Z}&0\to 0&\mathbb{Z}\to\mathbb{Z}_2&0\to\mathbb{Z}_2\\ 
&&n\mapsto 2n&&&&n\mapsto n&&n\mapsto n\\
\hline \hline
{\rm AI^\dag}& {\rm L} \to {\rm P}&\mathbb{Z} \to 0&0\to 0&0\to 0&0\to 2\mathbb{Z}&2\mathbb{Z}\to 0&0\to \mathbb{Z}_2&\mathbb{Z}_2\to \mathbb{Z}_2&\mathbb{Z}_2\to\mathbb{Z}\\ 
&&&&&&&&n\mapsto 0&\\
\hline
{\rm BDI^\dag}&{\rm L_r} \to {\rm P}&\mathbb{Z}_2 \to \mathbb{Z}&\mathbb{Z} \to 0&0\to 0&0\to 0&0\to 2\mathbb{Z}&2\mathbb{Z}\to 0&0\to \mathbb{Z}_2&\mathbb{Z}_2\to \mathbb{Z}_2\\ 
&&&&&&&&&n\mapsto 0\\ 
&{\rm L_i} \to {\rm P}&\mathbb{Z}\oplus \mathbb{Z}\to\mathbb{Z}&0\to 0&0\to 0&0\to 0&2\mathbb{Z}\oplus 2\mathbb{Z}\to 2\mathbb{Z}&0\to 0&\mathbb{Z}_2\oplus\mathbb{Z}_2\to\mathbb{Z}_2&\mathbb{Z}_2\oplus\mathbb{Z}_2\to\mathbb{Z}_2\\ 
&&(n,m)\mapsto n+m&&&&(n,m)\mapsto n+m&&(n,m)\mapsto n+m&(n,m)\mapsto n+m\\ 
\hline
{\rm D^\dag}&{\rm L_r} \to {\rm P}&\mathbb{Z}_2 \to \mathbb{Z}_2&\mathbb{Z}_2 \to \mathbb{Z}&\mathbb{Z} \to 0&0\to 0&0\to 0&0\to 2\mathbb{Z}&2\mathbb{Z}\to 0&0\to \mathbb{Z}_2\\ 
&&n \mapsto 0&&&&&&&\\
& {\rm L_i} \to {\rm P}&\mathbb{Z} \to \mathbb{Z}_2&0 \to \mathbb{Z}&0 \to 0&0 \to 0&2\mathbb{Z}\to 0&0\to 2\mathbb{Z}&\mathbb{Z}_2\to 0&\mathbb{Z}_2\to \mathbb{Z}_2 \\ 
&&n \mapsto n&&&&&&&n\mapsto n \\ 
\hline
{\rm DIII^\dag}& {\rm L_r} \to {\rm P}&0 \to \mathbb{Z}_2&\mathbb{Z}_2 \to \mathbb{Z}_2&\mathbb{Z}_2 \to \mathbb{Z}&\mathbb{Z} \to 0&0\to 0&0\to 0&0\to 2\mathbb{Z}&2\mathbb{Z}\to 0\\ 
&&&n \mapsto 0&&&&&&\\
&{\rm L_i} \to {\rm P}&\mathbb{Z}\to\mathbb{Z}_2&0\to\mathbb{Z}_2&\mathbb{Z}\to\mathbb{Z}&0\to 0&\mathbb{Z}\to 0&0\to 0&\mathbb{Z}\to 2\mathbb{Z}&0\to 0\\ 
&&n\mapsto n&&n\mapsto 2n&&&&n\mapsto n&\\ 
\hline
{\rm AII^\dag}& {\rm L} \to {\rm P}&2\mathbb{Z} \to 0&0 \to \mathbb{Z}_2&\mathbb{Z}_2 \to \mathbb{Z}_2&\mathbb{Z}_2 \to \mathbb{Z}&\mathbb{Z} \to 0&0\to 0&0\to 0&0\to 2\mathbb{Z}\\
&&&&n \mapsto 0&&&&&\\
\hline
{\rm CII^\dag}& {\rm L_r} \to {\rm P}&0 \to 2\mathbb{Z}&2\mathbb{Z} \to 0&0 \to \mathbb{Z}_2&\mathbb{Z}_2 \to \mathbb{Z}_2&\mathbb{Z}_2 \to \mathbb{Z}&\mathbb{Z} \to 0&0\to 0&0\to 0\\ 
&&&&&n \mapsto 0&&&&\\
&{\rm L_i} \to {\rm P}&2\mathbb{Z}\oplus 2\mathbb{Z}\to 2\mathbb{Z}&0\to 0&\mathbb{Z}_2\oplus\mathbb{Z}_2\to \mathbb{Z}_2&\mathbb{Z}_2\oplus \mathbb{Z}_2\to\mathbb{Z}_2&\mathbb{Z}\oplus \mathbb{Z}\to\mathbb{Z}&0\to 0&0\to 0&0\to 0\\ 
&&(n,m)\mapsto n+m&&(n,m)\mapsto n+m&(n,m)\mapsto n+m&(n,m)\mapsto n+m&&\\
\hline
{\rm C^\dag}& {\rm L_r} \to {\rm P}&0 \to 0&0 \to 2\mathbb{Z}&2\mathbb{Z} \to 0&0 \to \mathbb{Z}_2&\mathbb{Z}_2 \to \mathbb{Z}_2&\mathbb{Z}_2 \to \mathbb{Z}&\mathbb{Z} \to 0&0\to 0\\ 
&&&&&&n \mapsto 0&&&\\ 
& {\rm L_i} \to {\rm P}&2\mathbb{Z} \to 0&0\to 2\mathbb{Z}&\mathbb{Z}_2 \to 0&\mathbb{Z}_2 \to \mathbb{Z}_2&\mathbb{Z} \to \mathbb{Z}_2&0 \to \mathbb{Z}&0 \to 0&0 \to 0\\ 
&&&&&n \mapsto n&n \mapsto n&&&\\ 
\hline
{\rm CI^\dag}& {\rm L_r} \to {\rm P}&0 \to 0&0 \to 0&0 \to 2\mathbb{Z}&2\mathbb{Z} \to 0&0 \to \mathbb{Z}_2&\mathbb{Z}_2 \to \mathbb{Z}_2&\mathbb{Z}_2 \to \mathbb{Z}&\mathbb{Z} \to 0\\ 
&&&&&&&n \mapsto 0&&\\
& {\rm L_i} \to {\rm P}&\mathbb{Z}\to 0&0\to 0&\mathbb{Z}\to 2\mathbb{Z}&0\to 0&\mathbb{Z}\to\mathbb{Z}_2&0\to\mathbb{Z}_2&\mathbb{Z}\to\mathbb{Z}&0\to 0\\ 
&&&&n\mapsto n&&n\mapsto n&&n\mapsto 2n&\\
\hline \hline
\end{array}
$$
}
\end{sidewaystable}

\begin{sidewaystable}[]
\caption{Line gap homomorphisms for AZ class with an additional symmetry. }
\label{tab:SFH_AZ_add_2}
\centering
{\tiny
$$
\begin{array}{ccccccccccccc}
\mbox{AZ class} & \mbox{Add. symm.}& \mbox{Gap} &\delta=0&\delta=1&\delta=2&\delta=3&\delta=4&\delta=5&\delta=6&\delta=7 \\
\hline \hline
{\rm A}&\eta&{\rm L_r} \to {\rm P}&\mathbb{Z}\oplus\mathbb{Z}\to\mathbb{Z}&0\to 0&\mathbb{Z}\oplus\mathbb{Z}\to\mathbb{Z}&0\to 0&\mathbb{Z}\oplus\mathbb{Z}\to\mathbb{Z}&0\to 0&\mathbb{Z}\oplus\mathbb{Z}\to\mathbb{Z}&0\to 0\\
&&&(n,m)\mapsto n-m&&(n,m)\mapsto n-m&&(n,m)\mapsto n-m&&(n,m)\mapsto n-m&\\ 
&& {\rm L_i} \to {\rm P}&0 \to \mathbb{Z}&\mathbb{Z} \to 0&0 \to \mathbb{Z}&\mathbb{Z} \to 0&0 \to \mathbb{Z}&\mathbb{Z} \to 0&0 \to \mathbb{Z}&\mathbb{Z} \to 0\\
&&&&&&&&&&&\\ 
\hline
{\rm AIII}&S_+,\eta_{+}&{\rm L_r,L_i}\to {\rm P}&0\to 0&\mathbb{Z}\oplus\mathbb{Z}\to\mathbb{Z}&0\to 0&\mathbb{Z}\oplus\mathbb{Z}\to\mathbb{Z}&0\to 0&\mathbb{Z}\oplus\mathbb{Z}\to\mathbb{Z}&0\to 0&\mathbb{Z}\oplus\mathbb{Z}\to\mathbb{Z}\\ 
&&&&(n,m)\mapsto n-m&&(n,m)\mapsto n-m&&(n,m)\mapsto n-m&&(n,m)\mapsto n-m\\ 
\hline\hline
{\rm A}&S&{\rm L} \to {\rm P}&0\to 0&\mathbb{Z}\to\mathbb{Z}\oplus\mathbb{Z}&0\to 0&\mathbb{Z}\to\mathbb{Z}\oplus\mathbb{Z}&0\to 0&\mathbb{Z}\to\mathbb{Z}\oplus\mathbb{Z}&0\to 0&\mathbb{Z}\to\mathbb{Z}\oplus\mathbb{Z}\\ 
&&&&n\mapsto(n,n)&&n\mapsto(n,n)&&n\mapsto(n,n)&&n\mapsto(n,n)\\ 
\hline
{\rm AIII}&S_-,\eta_-&{\rm L_r\to P}&\mathbb{Z}\to\mathbb{Z}\oplus\mathbb{Z}&0\to 0&\mathbb{Z}\to\mathbb{Z}\oplus\mathbb{Z}&0\to 0&\mathbb{Z}\to\mathbb{Z}\oplus\mathbb{Z}&0\to 0&\mathbb{Z}\to\mathbb{Z}\oplus\mathbb{Z}&0\to 0\\
&&&n\mapsto(n,n)&&n\mapsto(n,n)&&n\mapsto(n,n)&&n\mapsto(n,n)&\\
&&{\rm L_i\to P}&\mathbb{Z}\to\mathbb{Z}\oplus\mathbb{Z}&0\to 0&\mathbb{Z}\to\mathbb{Z}\oplus\mathbb{Z}&0\to 0&\mathbb{Z}\to\mathbb{Z}\oplus\mathbb{Z}&0\to 0&\mathbb{Z}\to\mathbb{Z}\oplus\mathbb{Z}&0\to 0\\
&&&n\mapsto(n,-n)&&n\mapsto(n,-n)&&n\mapsto(n,-n)&&n\mapsto(n,-n)&\\
\hline\hline
{\rm AI}&\eta_+&{\rm L_r} \to {\rm P}&\mathbb{Z}\oplus \mathbb{Z}\to\mathbb{Z}&0\to 0&0\to 0&0\to 0&2\mathbb{Z}\oplus 2\mathbb{Z}\to 2\mathbb{Z}&0\to 0&\mathbb{Z}_2\oplus\mathbb{Z}_2\to\mathbb{Z}_2&\mathbb{Z}_2\oplus\mathbb{Z}_2\to\mathbb{Z}_2\\ 
&&&(n,m)\mapsto n+m&&&&(n,m)\mapsto n+m&&(n,m)\mapsto n+m&(n,m)\mapsto n+m\\ 
&&{\rm L_i} \to {\rm P}&\mathbb{Z}_2 \to \mathbb{Z}&\mathbb{Z} \to 0&0\to 0&0\to 0&0\to 2\mathbb{Z}&2\mathbb{Z}\to 0&0\to \mathbb{Z}_2&\mathbb{Z}_2\to \mathbb{Z}_2\\ 
&&&&&&&&&&n\mapsto 0\\ 
\hline
{\rm BDI}&S_{++},\eta_{++}&{\rm L_r,L_i}\to {\rm P}&\mathbb{Z}_2\oplus \mathbb{Z}_2\to\mathbb{Z}_2&\mathbb{Z}\oplus \mathbb{Z}\to\mathbb{Z}&0\to 0&0\to 0&0\to 0&2\mathbb{Z}\oplus 2\mathbb{Z}\to 2\mathbb{Z}&0\to 0&\mathbb{Z}_2\oplus\mathbb{Z}_2\to\mathbb{Z}_2\\ 
&&&(n,m)\mapsto n+m&(n,m)\mapsto n+m&&&&(n,m)\mapsto n+m&&(n,m)\mapsto n+m\\ 
\hline
{\rm D}&\eta_+& {\rm L_r} \to {\rm P}&\mathbb{Z}_2\oplus\mathbb{Z}_2\to \mathbb{Z}_2&\mathbb{Z}_2\oplus \mathbb{Z}_2\to\mathbb{Z}_2&\mathbb{Z}\oplus \mathbb{Z}\to\mathbb{Z}&0\to 0&0\to 0&0\to 0&2\mathbb{Z}\oplus 2\mathbb{Z}\to 2\mathbb{Z}&0\to 0\\ 
&&&(n,m)\mapsto n+m&(n,m)\mapsto n+m&(n,m)\mapsto n+m&&&&(n,m)\mapsto n+m&\\
&& {\rm L_i\to P}&\mathbb{Z}_2 \to \mathbb{Z}_2&\mathbb{Z} \to \mathbb{Z}_2&0 \to \mathbb{Z}&0 \to 0&0 \to 0&2\mathbb{Z}\to 0&0\to 2\mathbb{Z}&\mathbb{Z}_2\to 0\\ 
&&&n \mapsto n&n \mapsto n&&&&&&\\ 
\hline
{\rm DIII}&S_{--},\eta_{++}& {\rm L_r}\to {\rm P}&0\to 0&\mathbb{Z}_2\oplus\mathbb{Z}_2\to \mathbb{Z}_2&\mathbb{Z}_2\oplus \mathbb{Z}_2\to\mathbb{Z}_2&\mathbb{Z}\oplus \mathbb{Z}\to\mathbb{Z}&0\to 0&0\to 0&0\to 0&2\mathbb{Z}\oplus 2\mathbb{Z}\to 2\mathbb{Z}\\ 
&&&&(n,m)\mapsto n+m&(n,m)\mapsto n+m&(n,m)\mapsto n+m&&&&(n,m)\mapsto n+m\\ 
&&{\rm L_i}\to {\rm P}&0\to 0&\mathbb{Z}\to\mathbb{Z}_2&0\to\mathbb{Z}_2&\mathbb{Z}\to\mathbb{Z}&0\to 0&\mathbb{Z}\to 0&0\to 0&\mathbb{Z}\to 2\mathbb{Z}\\ 
&&&&n\mapsto n&&n\mapsto 2n&&&&n\mapsto n\\ 
\hline
{\rm AII}&\eta_+& {\rm L_r} \to {\rm P}&2\mathbb{Z}\oplus 2\mathbb{Z}\to 2\mathbb{Z}&0\to 0&\mathbb{Z}_2\oplus\mathbb{Z}_2\to \mathbb{Z}_2&\mathbb{Z}_2\oplus \mathbb{Z}_2\to\mathbb{Z}_2&\mathbb{Z}\oplus \mathbb{Z}\to\mathbb{Z}&0\to 0&0\to 0&0\to 0\\ 
&&&(n,m)\mapsto n+m&&(n,m)\mapsto n+m&(n,m)\mapsto n+m&(n,m)\mapsto n+m&&\\
&&{\rm L_i} \to {\rm P}&0 \to 2\mathbb{Z}&2\mathbb{Z} \to 0&0 \to \mathbb{Z}_2&\mathbb{Z}_2 \to \mathbb{Z}_2&\mathbb{Z}_2 \to \mathbb{Z}&\mathbb{Z} \to 0&0\to 0&0\to 0\\ 
&&&&&&n \mapsto 0&&&&\\
\hline
{\rm CII}&S_{++},\eta_{++}& {\rm L_r,L_i} \to {\rm P}&0\to 0&2\mathbb{Z}\oplus 2\mathbb{Z}\to 2\mathbb{Z}&0\to 0&\mathbb{Z}_2\oplus\mathbb{Z}_2\to \mathbb{Z}_2&\mathbb{Z}_2\oplus \mathbb{Z}_2\to\mathbb{Z}_2&\mathbb{Z}\oplus \mathbb{Z}\to\mathbb{Z}&0\to 0&0\to 0\\ 
&&&&(n,m)\mapsto n+m&&(n,m)\mapsto n+m&(n,m)\mapsto n+m&(n,m)\mapsto n+m&\\ 
\hline
{\rm C}&\eta_+& {\rm L_r} \to {\rm P}&0\to 0&0\to 0&2\mathbb{Z}\oplus 2\mathbb{Z}\to 2\mathbb{Z}&0\to 0&\mathbb{Z}_2\oplus\mathbb{Z}_2\to \mathbb{Z}_2&\mathbb{Z}_2\oplus \mathbb{Z}_2\to\mathbb{Z}_2&\mathbb{Z}\oplus \mathbb{Z}\to\mathbb{Z}&0\to 0\\ 
&&&&&(n,m)\mapsto n+m&&(n,m)\mapsto n+m&(n,m)\mapsto n+m&(n,m)\mapsto n+m\\
&& {\rm L_i} \to {\rm P}&0 \to 0&2\mathbb{Z} \to 0&0\to 2\mathbb{Z}&\mathbb{Z}_2 \to 0&\mathbb{Z}_2 \to \mathbb{Z}_2&\mathbb{Z} \to \mathbb{Z}_2&0 \to \mathbb{Z}&0 \to 0\\ 
&&&&&&&n \mapsto n&n \mapsto n&&\\ 
\hline
{\rm CI}&S_{--},\eta_{++}& {\rm L_r} \to {\rm P}&0\to 0&0\to 0&0\to 0&2\mathbb{Z}\oplus 2\mathbb{Z}\to 2\mathbb{Z}&0\to 0&\mathbb{Z}_2\oplus\mathbb{Z}_2\to \mathbb{Z}_2&\mathbb{Z}_2\oplus \mathbb{Z}_2\to\mathbb{Z}_2&\mathbb{Z}\oplus \mathbb{Z}\to\mathbb{Z}\\ 
&&&&&&(n,m)\mapsto n+m&&(n,m)\mapsto n+m&(n,m)\mapsto n+m&(n,m)\mapsto n+m\\ 
&& {\rm L_i} \to {\rm P}&0\to 0&\mathbb{Z}\to 0&0\to 0&\mathbb{Z}\to 2\mathbb{Z}&0\to 0&\mathbb{Z}\to\mathbb{Z}_2&0\to\mathbb{Z}_2&\mathbb{Z}\to\mathbb{Z}\\ 
&&&&&&n\mapsto n&&n\mapsto n&&n\mapsto 2n\\ 
\hline
\hline
{\rm AI}&S_-&{\rm L_r \to \rm P}&0\to 0&0\to\mathbb{Z}&0\to 0&2\mathbb{Z}\to\mathbb{Z}&0\to 0&\mathbb{Z}_2\to\mathbb{Z}&\mathbb{Z}_2\to 0&\mathbb{Z}\to\mathbb{Z}\\
&&&&&&n\mapsto 2n&&&&n\mapsto n\\
&&{\rm L_i \to \rm P}&0\to 0&\mathbb{Z}_2\to \mathbb{Z}&\mathbb{Z}_2\to 0&\mathbb{Z}\to \mathbb{Z}&0\to 0&0\to\mathbb{Z}&0\to 0&2\mathbb{Z}\to\mathbb{Z}\\
&&&&&&n\mapsto n&&&&n\mapsto 2n\\
\hline
{\rm BDI}&S_{-+},\eta_{+-}&{\rm L_r \to \rm P}&\mathbb{Z}\to \mathbb{Z}&0\to 0&0\to\mathbb{Z}&0\to 0&2\mathbb{Z}\to\mathbb{Z}&0\to 0&\mathbb{Z}_2\to\mathbb{Z}&\mathbb{Z}_2\to 0\\
&&&n\mapsto n&&&&n\mapsto 2n&&&\\
&&{\rm L_i \to \rm P}&\mathbb{Z}_2\to \mathbb{Z}&\mathbb{Z}_2\to 0&\mathbb{Z}\to \mathbb{Z}&0\to 0&0\to\mathbb{Z}&0\to 0&2\mathbb{Z}\to\mathbb{Z}&0\to 0\\
&&&&&n\mapsto n&&&&n\mapsto 2n&\\
\hline
{\rm D}&S_+&{\rm L \to \rm P}&\mathbb{Z}_2\to 0&\mathbb{Z}\to \mathbb{Z}&0\to 0&0\to\mathbb{Z}&0\to 0&2\mathbb{Z}\to\mathbb{Z}&0\to 0&\mathbb{Z}_2\to\mathbb{Z}\\
&&&&n\mapsto n&&&&n\mapsto 2n&&\\
\hline
{\rm DIII}&S_{-+},\eta_{-+}&{\rm L_r \to \rm P}&\mathbb{Z}_2\to \mathbb{Z}&\mathbb{Z}_2\to 0&\mathbb{Z}\to \mathbb{Z}&0\to 0&0\to\mathbb{Z}&0\to 0&2\mathbb{Z}\to\mathbb{Z}&0\to 0\\
&&&&&n\mapsto n&&&&n\mapsto 2n&\\
&&{\rm L_i \to \rm P}&\mathbb{Z}\to \mathbb{Z}&0\to 0&0\to\mathbb{Z}&0\to 0&2\mathbb{Z}\to\mathbb{Z}&0\to 0&\mathbb{Z}_2\to\mathbb{Z}&\mathbb{Z}_2\to 0\\
&&&n\mapsto n&&&&n\mapsto 2n&&&\\
\hline
{\rm AII}&S_-&{\rm L_r \to \rm P}&0\to 0&\mathbb{Z}_2\to \mathbb{Z}&\mathbb{Z}_2\to 0&\mathbb{Z}\to \mathbb{Z}&0\to 0&0\to\mathbb{Z}&0\to 0&2\mathbb{Z}\to\mathbb{Z}\\
&&&&&&n\mapsto n&&&&n\mapsto 2n\\
&&{\rm L_i \to \rm P}&0\to 0&0\to\mathbb{Z}&0\to 0&2\mathbb{Z}\to\mathbb{Z}&0\to 0&\mathbb{Z}_2\to\mathbb{Z}&\mathbb{Z}_2\to 0&\mathbb{Z}\to\mathbb{Z}\\
&&&&&&n\mapsto 2n&&&&n\mapsto n\\
\hline
{\rm CII}&S_{-+},\eta_{+-}&{\rm L_r \to \rm P}&2\mathbb{Z}\to \mathbb{Z}&0\to 0&\mathbb{Z}_2\to \mathbb{Z}&\mathbb{Z}_2\to 0&\mathbb{Z}\to \mathbb{Z}&0\to 0&0\to\mathbb{Z}&0\to 0\\
&&&n\mapsto 2n&&&&n\mapsto n&&&\\
&&{\rm L_i \to \rm P}&0\to \mathbb{Z}&0\to 0&2\mathbb{Z}\to \mathbb{Z}&0\to 0&\mathbb{Z}_2\to \mathbb{Z}&\mathbb{Z}_2\to 0&\mathbb{Z}\to \mathbb{Z}&0\to 0\\
&&&&&n\mapsto 2n&&&&n\mapsto n&\\
\hline
{\rm C}&S_+&{\rm L \to \rm P}&0\to 0&2\mathbb{Z}\to \mathbb{Z}&0\to 0&\mathbb{Z}_2\to \mathbb{Z}&\mathbb{Z}_2\to 0&\mathbb{Z}\to \mathbb{Z}&0\to 0&0\to\mathbb{Z}\\
&&&&n\mapsto 2n&&&&n\mapsto n&&\\
\hline 
{\rm CI}&S_{-+},\eta_{-+}&{\rm L_r \to \rm P}&0\to \mathbb{Z}&0\to 0&2\mathbb{Z}\to \mathbb{Z}&0\to 0&\mathbb{Z}_2\to \mathbb{Z}&\mathbb{Z}_2\to 0&\mathbb{Z}\to \mathbb{Z}&0\to 0\\
&&&&&n\mapsto 2n&&&&n\mapsto n&\\
&&{\rm L_i \to \rm P}&2\mathbb{Z}\to \mathbb{Z}&0\to 0&\mathbb{Z}_2\to \mathbb{Z}&\mathbb{Z}_2\to 0&\mathbb{Z}\to \mathbb{Z}&0\to 0&0\to\mathbb{Z}&0\to 0\\
&&&n\mapsto 2n&&&&n\mapsto n&&&\\
\hline \hline
\end{array}
$$
}
\end{sidewaystable}

\begin{sidewaystable*}[!]
\caption*{Continuation of Table~\ref{tab:SFH_AZ_add_2}.}
\centering
{\tiny
$$
\begin{array}{ccccccccccccc}
\mbox{Symm. class} & \mbox{Add. symm.}& \mbox{Gap} &\delta=0&\delta=1&\delta=2&\delta=3&\delta=4&\delta=5&\delta=6&\delta=7 \\
\hline \hline
{\rm AI}&\eta_-&{\rm L_r} \to {\rm P}&\mathbb{Z}\to\mathbb{Z}_2&0\to\mathbb{Z}_2&\mathbb{Z}\to\mathbb{Z}&0\to 0&\mathbb{Z}\to 0&0\to 0&\mathbb{Z}\to 2\mathbb{Z}&0\to 0\\ 
&&&n\mapsto n&&n\mapsto 2n&&&&n\mapsto n&\\ 
&& {\rm L_i} \to {\rm P}&0 \to \mathbb{Z}_2&\mathbb{Z}_2 \to \mathbb{Z}_2&\mathbb{Z}_2 \to \mathbb{Z}&\mathbb{Z} \to 0&0\to 0&0\to 0&0\to 2\mathbb{Z}&2\mathbb{Z}\to 0\\ 
&&&&n \mapsto 0&&&&&&\\
\hline
{\rm BDI}&S_{--},\eta_{--}&{\rm L_r}\to {\rm P}&0\to 0&\mathbb{Z}\to\mathbb{Z}_2&0\to\mathbb{Z}_2&\mathbb{Z}\to\mathbb{Z}&0\to 0&\mathbb{Z}\to 0&0\to 0&\mathbb{Z}\to 2\mathbb{Z}\\ 
&&&&n\mapsto n&&n\mapsto 2n&&&&n\mapsto n\\ 
&& {\rm L_i}\to {\rm P}&0\to 0&\mathbb{Z}_2\oplus\mathbb{Z}_2\to \mathbb{Z}_2&\mathbb{Z}_2\oplus \mathbb{Z}_2\to\mathbb{Z}_2&\mathbb{Z}\oplus \mathbb{Z}\to\mathbb{Z}&0\to 0&0\to 0&0\to 0&2\mathbb{Z}\oplus 2\mathbb{Z}\to 2\mathbb{Z}\\ 
&&&&(n,m)\mapsto n+m&(n,m)\mapsto n+m&(n,m)\mapsto n+m&&&&(n,m)\mapsto n+m\\ 
\hline
{\rm D}&\eta_-& {\rm L_r} \to {\rm P}&\mathbb{Z}\to 2\mathbb{Z}&0\to 0&\mathbb{Z}\to\mathbb{Z}_2&0\to\mathbb{Z}_2&\mathbb{Z}\to\mathbb{Z}&0\to 0&\mathbb{Z}\to 0&0\to 0\\ 
&&&n\mapsto n&&n\mapsto n&&n\mapsto 2n&&\\
&&{\rm L_i} \to {\rm P}&0\to 2\mathbb{Z}&\mathbb{Z}_2 \to 0&\mathbb{Z}_2 \to \mathbb{Z}_2&\mathbb{Z} \to \mathbb{Z}_2&0 \to \mathbb{Z}&0 \to 0&0 \to 0&2\mathbb{Z}\to 0\\ 
&&&&&n \mapsto n&n \mapsto n&&&&\\ 
\hline
{\rm DIII}&S_{++},\eta_{--}& {\rm L_r,L_i}\to {\rm P}&0\to 0&\mathbb{Z}\to 2\mathbb{Z}&0\to 0&\mathbb{Z}\to\mathbb{Z}_2&0\to\mathbb{Z}_2&\mathbb{Z}\to\mathbb{Z}&0\to 0&\mathbb{Z}\to 0\\ 
&&&&n\mapsto n&&n\mapsto n&&n\mapsto 2n&&\\ 
\hline
{\rm AII}&\eta_-& {\rm L_r} \to {\rm P}&\mathbb{Z}\to 0&0\to 0&\mathbb{Z}\to 2\mathbb{Z}&0\to 0&\mathbb{Z}\to\mathbb{Z}_2&0\to\mathbb{Z}_2&\mathbb{Z}\to\mathbb{Z}&0\to 0\\ 
&&&&&n\mapsto n&&n\mapsto n&&n\mapsto 2n&\\
&& {\rm L_i} \to {\rm P}&0 \to 0&0 \to 0&0 \to 2\mathbb{Z}&2\mathbb{Z} \to 0&0 \to \mathbb{Z}_2&\mathbb{Z}_2 \to \mathbb{Z}_2&\mathbb{Z}_2 \to \mathbb{Z}&\mathbb{Z} \to 0\\ 
&&&&&&&&n \mapsto 0&&\\
\hline 
{\rm CII}&S_{--},\eta_{--}& {\rm L_r} \to {\rm P}&0\to 0&\mathbb{Z}\to 0&0\to 0&\mathbb{Z}\to 2\mathbb{Z}&0\to 0&\mathbb{Z}\to\mathbb{Z}_2&0\to\mathbb{Z}_2&\mathbb{Z}\to\mathbb{Z}\\ 
&&&&&&n\mapsto n&&n\mapsto n&&n\mapsto 2n\\ 
&& {\rm L_i} \to {\rm P}&0\to 0&0\to 0&0\to 0&2\mathbb{Z}\oplus 2\mathbb{Z}\to 2\mathbb{Z}&0\to 0&\mathbb{Z}_2\oplus\mathbb{Z}_2\to \mathbb{Z}_2&\mathbb{Z}_2\oplus \mathbb{Z}_2\to\mathbb{Z}_2&\mathbb{Z}\oplus \mathbb{Z}\to\mathbb{Z}\\ 
&&&&&&(n,m)\mapsto n+m&&(n,m)\mapsto n+m&(n,m)\mapsto n+m&(n,m)\mapsto n+m\\ 
\hline
{\rm C}&\eta_-& {\rm L_r} \to {\rm P}&\mathbb{Z}\to \mathbb{Z}&0\to 0&\mathbb{Z}\to 0&0\to 0&\mathbb{Z}\to 2\mathbb{Z}&0\to 0&\mathbb{Z}\to\mathbb{Z}_2&0\to\mathbb{Z}_2\\ 
&&&n\mapsto 2n&&&&n\mapsto n&&n\mapsto n\\
&& {\rm L_i} \to {\rm P}&0 \to \mathbb{Z}&0 \to 0&0 \to 0&2\mathbb{Z} \to 0&0\to 2\mathbb{Z}&\mathbb{Z}_2 \to 0&\mathbb{Z}_2 \to \mathbb{Z}_2&\mathbb{Z} \to \mathbb{Z}_2\\ 
&&&&&&&&&n \mapsto n&n \mapsto n\\ 
\hline 
{\rm CI}&S_{++},\eta_{--}& {\rm L_r,L_i} \to {\rm P}&0\to \mathbb{Z}_2&\mathbb{Z}\to \mathbb{Z}&0\to 0&\mathbb{Z}\to 0&0\to 0&\mathbb{Z}\to 2\mathbb{Z}&0\to 0&\mathbb{Z}\to\mathbb{Z}_2\\ 
&&&&n\mapsto 2n&&&&n\mapsto n&&n\mapsto n\\ 
\hline \hline
{\rm AI}&S_+&{\rm L_r \to \rm P}&\mathbb{Z}_2\to\mathbb{Z}_2\oplus\mathbb{Z}_2&\mathbb{Z}\to\mathbb{Z}\oplus\mathbb{Z}&0\to 0&0\to 0&0\to 0&\mathbb{Z}\to\mathbb{Z}\oplus\mathbb{Z}&0\to 0&\mathbb{Z}_2\to\mathbb{Z}_2\oplus\mathbb{Z}_2\\
&&&n\mapsto(n,n)&n\mapsto(n,n)&&&&n\mapsto(n,n)&&n\mapsto(n,n)\\
&&{\rm L_i \to \rm P}&\mathbb{Z}_2\to\mathbb{Z}_2\oplus\mathbb{Z}_2&\mathbb{Z}\to\mathbb{Z}\oplus\mathbb{Z}&0\to 0&0\to 0&0\to 0&\mathbb{Z}\to\mathbb{Z}\oplus\mathbb{Z}&0\to 0&\mathbb{Z}_2\to\mathbb{Z}_2\oplus\mathbb{Z}_2\\
&&&n\mapsto(n,n)&n\mapsto(n,n)&&&&n\mapsto(n,n)&&n\mapsto(n,n)\\
\hline
{\rm BDI}&S_{+-},\eta_{-+}&{\rm L_r \to \rm P}&\mathbb{Z}_2\to\mathbb{Z}_2\oplus\mathbb{Z}_2&\mathbb{Z}_2\to\mathbb{Z}_2\oplus\mathbb{Z}_2&\mathbb{Z}\to\mathbb{Z}\oplus\mathbb{Z}&0\to 0&0\to 0&0\to 0&\mathbb{Z}\to\mathbb{Z}\oplus\mathbb{Z}&0\to 0\\
&&&n\mapsto(n,n)&n\mapsto(n,n)&n\mapsto(n,n)&&&&n\mapsto(n,n)&\\
&&{\rm L_i \to \rm P}&\mathbb{Z}_2\to\mathbb{Z}_2\oplus\mathbb{Z}_2&\mathbb{Z}_2\to\mathbb{Z}_2\oplus\mathbb{Z}_2&\mathbb{Z}\to\mathbb{Z}\oplus\mathbb{Z}&0\to 0&0\to 0&0\to 0&\mathbb{Z}\to\mathbb{Z}\oplus\mathbb{Z}&0\to 0\\
&&&n\mapsto(n,n)&n\mapsto(n,n)&n\mapsto(n,-n)&&&&n\mapsto(n,-n)&\\
\hline
{\rm D}&S_-&{\rm L \to \rm P}&0\to 0&\mathbb{Z}_2\to\mathbb{Z}_2\oplus\mathbb{Z}_2&\mathbb{Z}_2\to\mathbb{Z}_2\oplus\mathbb{Z}_2&\mathbb{Z}\to\mathbb{Z}\oplus\mathbb{Z}&0\to 0&0\to 0&0\to 0&\mathbb{Z}\to\mathbb{Z}\oplus\mathbb{Z}\\
&&&&n\mapsto(n,n)&n\mapsto(n,n)&n\mapsto(n,n)&&&&n\mapsto(n,n)\\
\hline
{\rm DIII}&S_{+-},\eta_{+-}&{\rm L_r \to \rm P}&\mathbb{Z}\to\mathbb{Z}\oplus\mathbb{Z}&0\to 0&\mathbb{Z}_2\to\mathbb{Z}_2\oplus\mathbb{Z}_2&\mathbb{Z}_2\to\mathbb{Z}_2\oplus\mathbb{Z}_2&\mathbb{Z}\to\mathbb{Z}\oplus\mathbb{Z}&0\to 0&0\to 0&0\to 0\\
&&&n\mapsto(n,n)&&n\mapsto(n,n)&n\mapsto(n,n)&n\mapsto(n,n)&&&\\
&&{\rm L_i \to \rm P}&\mathbb{Z}\to\mathbb{Z}\oplus\mathbb{Z}&0\to 0&\mathbb{Z}_2\to\mathbb{Z}_2\oplus\mathbb{Z}_2&\mathbb{Z}_2\to\mathbb{Z}_2\oplus\mathbb{Z}_2&\mathbb{Z}\to\mathbb{Z}\oplus\mathbb{Z}&0\to 0&0\to 0&0\to 0\\
&&&n\mapsto(n,-n)&&n\mapsto(n,n)&n\mapsto(n,n)&n\mapsto(n,-n)&&&\\
\hline
{\rm AII}&S_+&{\rm L_r\to \rm P}&0\to 0&\mathbb{Z}\to\mathbb{Z}\oplus\mathbb{Z}&0\to 0&\mathbb{Z}_2\to\mathbb{Z}_2\oplus\mathbb{Z}_2&\mathbb{Z}_2\to\mathbb{Z}_2\oplus\mathbb{Z}_2&\mathbb{Z}\to\mathbb{Z}\oplus\mathbb{Z}&0\to 0&0\to 0\\
&&&&n\mapsto(n,n)&&n\mapsto(n,n)&n\mapsto(n,n)&n\mapsto(n,n)&&\\
&&{\rm L_i \to \rm P}&0\to 0&\mathbb{Z}\to\mathbb{Z}\oplus\mathbb{Z}&0\to 0&\mathbb{Z}_2\to\mathbb{Z}_2\oplus\mathbb{Z}_2&\mathbb{Z}_2\to\mathbb{Z}_2\oplus\mathbb{Z}_2&\mathbb{Z}\to\mathbb{Z}\oplus\mathbb{Z}&0\to 0&0\to 0\\
&&&&n\mapsto(n,n)&&n\mapsto(n,n)&n\mapsto(n,n)&n\mapsto(n,n)&&\\
\hline
{\rm CII}&S_{+-},\eta{-+}&{\rm L_r \to \rm P}&0\to 0&0\to 0&\mathbb{Z}\to\mathbb{Z}\oplus\mathbb{Z}&0\to 0&\mathbb{Z}_2\to\mathbb{Z}_2\oplus\mathbb{Z}_2&\mathbb{Z}_2\to\mathbb{Z}_2\oplus\mathbb{Z}_2&\mathbb{Z}\to\mathbb{Z}\oplus\mathbb{Z}&0\to 0\\
&&&&&n\mapsto(n,n)&&n\mapsto(n,n)&n\mapsto(n,n)&n\mapsto(n,n)&\\
&&{\rm L_i \to \rm P}&0\to 0&0\to 0&\mathbb{Z}\to\mathbb{Z}\oplus\mathbb{Z}&0\to 0&\mathbb{Z}_2\to\mathbb{Z}_2\oplus\mathbb{Z}_2&\mathbb{Z}_2\to\mathbb{Z}_2\oplus\mathbb{Z}_2&\mathbb{Z}\to\mathbb{Z}\oplus\mathbb{Z}&0\to 0\\
&&&&&n\mapsto(n,-n)&&n\mapsto(n,n)&n\mapsto(n,n)&n\mapsto(n,-n)&\\
\hline
{\rm C}&S_-&{\rm L \to \rm P}&0\to 0&0\to 0&0\to 0&\mathbb{Z}\to\mathbb{Z}\oplus\mathbb{Z}&0\to 0&\mathbb{Z}_2\to\mathbb{Z}_2\oplus\mathbb{Z}_2&\mathbb{Z}_2\to\mathbb{Z}_2\oplus\mathbb{Z}_2&\mathbb{Z}\to\mathbb{Z}\oplus\mathbb{Z}\\
&&&&&&n\mapsto(n,n)&&n\mapsto(n,n)&n\mapsto(n,n)&n\mapsto(n,n)\\
\hline
{\rm CI}&S_{+-},\eta_{+-}&{\rm L_r \to \rm P}&\mathbb{Z}\to\mathbb{Z}\oplus\mathbb{Z}&0\to 0&0\to 0&0\to 0&\mathbb{Z}\to\mathbb{Z}\oplus\mathbb{Z}&0\to 0&\mathbb{Z}_2\to\mathbb{Z}_2\oplus\mathbb{Z}_2&\mathbb{Z}_2\to\mathbb{Z}_2\oplus\mathbb{Z}_2\\
&&&n\mapsto(n,n)&&&&n\mapsto(n,n)&&n\mapsto(n,n)&n\mapsto(n,n)\\
&&{\rm L_i \to \rm P}&\mathbb{Z}\to\mathbb{Z}\oplus\mathbb{Z}&0\to 0&0\to 0&0\to 0&\mathbb{Z}\to\mathbb{Z}\oplus\mathbb{Z}&0\to 0&\mathbb{Z}_2\to\mathbb{Z}_2\oplus\mathbb{Z}_2&\mathbb{Z}_2\to\mathbb{Z}_2\oplus\mathbb{Z}_2\\
&&&n\mapsto(n,-n)&&&&n\mapsto(n,-n)&&n\mapsto(n,n)&n\mapsto(n,n)\\
\hline \hline
\end{array}
$$
}
\end{sidewaystable*}

\begin{table}[]
\caption{The classification table of intrinsic point-gap topology for AZ and AZ$^\dag$ classes.}
\label{tab:SFH_quotient_AZ}
\centering
$$
\begin{array}{ccccccccccccc}
\mbox{AZ class}&\delta=0&\delta=1&\delta=2&\delta=3&\delta=4&\delta=5&\delta=6&\delta=7 \\
\hline \hline
{\rm A}&0&\Z&0&\Z&0&\Z&0&\Z\\ 
{\rm AIII}&0&0&0&0&0&0&0&0\\
{\rm AI}&0&\Z&0&0&0&2\Z&0&0\\
{\rm BDI}&0&0&0&0&0&0&0&0\\
{\rm D}&0&0&0&\Z&0&0&0&2\Z\\
{\rm DIII}&0&0&0&0&\Z_2&0&0&0\\
{\rm AII}&0&2\Z&0&0&0&\Z&0&0\\
{\rm CII}&0&0&0&0&0&0&0&0\\
{\rm C}&0&0&0&2\Z&0&0&0&\Z\\
{\rm CI}&\Z_2&0&0&0&0&0&0&0\\
\hline
{\rm AI^\dag}&0&0&0&2\Z&0&\Z_2&\Z_2&\Z\\
{\rm BDI^\dag}&0&0&0&0&0&0&0&0\\ 
{\rm D^\dag}&0&\Z&0&0&0&2\Z&0&0\\
{\rm DIII^\dag}&0&\Z_2&\Z_2&0&0&0&0&0\\ 
{\rm AII^\dag}&0&\Z_2&\Z_2&\Z&0&0&0&2\Z\\
{\rm CII^\dag}&0&0&0&0&0&0&0&0\\ 
{\rm C^\dag}&0&2\Z&0&0&0&\Z&0&0\\
{\rm CI^\dag}&0&0&0&0&0&\Z_2&\Z_2&0\\
\hline
\hline
\end{array}
$$
\end{table}

\begin{table}[]
\caption{The classification table of intrinsic point-gap topology for AZ class with sublattice symmetry or pseudo-Hermiticity. }
\label{tab:SFH_quotient_AZ_add}
\centering
$$
\begin{array}{ccccccccccccc}
\mbox{AZ class}&\mbox{Add. symm.}&\delta=0&\delta=1&\delta=2&\delta=3&\delta=4&\delta=5&\delta=6&\delta=7 \\
\hline \hline
{\rm A}&\eta&0&0&0&0&0&0&0&0\\
{\rm AIII}&S_+,\eta_+&0&0&0&0&0&0&0&0\\
\hline 
{\rm A}&S&0&\Z&0&\Z&0&\Z&0&\Z\\ 
{\rm AIII}&S_-,\eta_-&\Z_2&0&\Z_2&0&\Z_2&0&\Z_2&0\\
\hline
{\rm AI}&\eta_+&0&0&0&0&0&0&0&0\\ 
{\rm BDI}&S_{++},\eta_{++}&0&0&0&0&0&0&0&0\\ 
{\rm D}&\eta_+&0&0&0&0&0&0&0&0\\ 
{\rm DIII}&S_{--},\eta_{++}&0&0&0&0&0&0&0&0\\ 
{\rm AII}&\eta_+&0&0&0&0&0&0&0&0\\ 
{\rm CII}&S_{++},\eta_{++}&0&0&0&0&0&0&0&0\\ 
{\rm C}&\eta_+&0&0&0&0&0&0&0&0\\ 
{\rm CI}&S_{--},\eta_{++}&0&0&0&0&0&0&0&0\\ 
\hline
{\rm AI}&S_-&0&\Z&0&0&0&\Z&0&0\\
{\rm BDI}&S_{-+},\eta_{+-}&0&0&0&0&\Z_2&0&\Z_2&0\\
{\rm D}&S_+&0&0&0&\Z&0&\Z_2&0&\Z\\
{\rm DIII}&S_{-+},\eta_{-+}&0&0&0&0&\Z_2&0&\Z_2&0\\
{\rm AII}&S_-&0&\Z&0&0&0&\Z&0&0&\\
{\rm CII}&S_{-+},\eta_{+-}&\Z_2&0&\Z_2&0&0&0&0&0\\
{\rm C}&S_+&0&\Z_2&0&\Z&0&0&0&\Z\\
{\rm CI}&S_{-+},\eta_{-+}&\Z_2&0&\Z_2&0&0&0&0&0\\
\hline
{\rm AI}&\eta_-&0&\Z_2&\Z_2&0&0&0&0&0\\
{\rm BDI}&S_{--},\eta_{--}&0&0&0&0&0&0&0&0\\ 
{\rm D}&\eta_-&0&0&0&0&\Z_2&0&0&0\\ 
{\rm DIII}&S_{++},\eta_{--}&0&0&0&0&\Z_2&\Z_2&0&0\\ 
{\rm AII}&\eta_-&0&0&0&0&0&\Z_2&\Z_2&0\\ 
{\rm CII}&S_{--},\eta_{--}&0&0&0&0&0&0&0&0\\ 
{\rm C}&\eta_-&\Z_2&0&0&0&0&0&0&0\\ 
{\rm CI}&S_{++},\eta_{--}&\Z_2&\Z_2&0&0&0&0&0&0\\ 
\hline
{\rm AI}&S_+&\Z_2&\Z&0&0&0&\Z&0&\Z_2\\
{\rm BDI}&S_{+-},\eta_{-+}&\Z_2&\Z_2&\Z_2&0&0&0&\Z_2&0\\
{\rm D}&S_-&0&\Z_2&\Z_2&\Z&0&0&0&\Z\\
{\rm DIII}&S_{+-},\eta_{+-}&\Z_2&0&\Z_2&\Z_2&\Z_2&0&0&0\\
{\rm AII}&S_+&0&\Z&0&\Z_2&\Z_2&\Z&0&0\\
{\rm CII}&S_{+-},\eta{-+}&0&0&\Z_2&0&\Z_2&\Z_2&\Z_2&0\\
{\rm C}&S_-&0&0&0&\Z&0&\Z_2&\Z_2&\Z\\
{\rm CI}&S_{+-},\eta_{+-}&\Z_2&0&0&0&\Z_2&0&\Z_2&\Z_2\\
\hline \hline
\end{array}
$$
\end{table}

\section{Summary and Outlook}

In this work, we have formulated a general framework for classifying non-Hermitian topological phases.  
By defining homomorphisms from real- and imaginary-line-gap phases to point-gap phases, we identified 
intrinsic non-Hermitian topological phases as those that cannot be reduced to Hermitian counterparts.  
We computed these homomorphisms explicitly for all 54 internal symmetry classes. The resulting tables provide a systematic account of intrinsic and extrinsic phases.

The present formulation establishes the algebraic structure that underlies intrinsic non-Hermitian 
topology and clarifies its relation to Hermitian classifications. The results show that intrinsic phases 
arise in a broad range of symmetry settings and dimensions, and include phenomena without Hermitian 
analogs such as the non-Hermitian skin effect.

Several directions remain for future work. One is to extend the present framework to systems with 
crystalline symmetries, where spatial symmetries impose further constraints on possible intrinsic phases.  
Another is to investigate the interplay with interacting systems and stability beyond free-fermion models.  
It is also of interest to study concrete physical realizations of the intrinsic phases identified here, 
in photonic, acoustic, or other experimental platforms. These problems go beyond the present work and 
are left for future studies.

\section*{Acknowledgements}
We were supported by JST CREST Grant No. JPMJCR19T2, and JSPS KAKENHI Grant No. JP22H05118 and JP23H01097.

\bibliography{refs}

\end{document}